\documentclass[a4paper,fleqn]{cas-dc}

\usepackage{bm}
\usepackage[numbers]{natbib}
\usepackage{empheq}
\newcommand{\sign}{\text{sign}}
\newtheorem{assumption}{Assumption}
\newtheorem{remark}{Remark}
\newtheorem{notation}{Notation}
\newtheorem{definition}{Definition}
\newtheorem{proposition}{Proposition}
\newtheorem{theorem}{Theorem}

\newproof{pf}{Proof}
\newcommand{\tabincell}[2]{\begin{tabular}{@{}#1@{}}#2\end{tabular}}
\usepackage{ragged2e}

\def\tsc#1{\csdef{#1}{\textsc{\lowercase{#1}}\xspace}}
\tsc{WGM}
\tsc{QE}
\tsc{EP}
\tsc{PMS}
\tsc{BEC}
\tsc{DE}

\begin{document}
\let\WriteBookmarks\relax
\def\floatpagepagefraction{1}
\def\textpagefraction{.001}
\shorttitle{Title:} 
\shortauthors{Author 1 et~al.}

\title [mode = title]{Adaptive dynamic programming-based adaptive-gain sliding mode tracking control for fixed-wing UAV with disturbances}

\author[1]{Chaofan Zhang}[style=chinese]

\author[1]{Guoshan Zhang}[style=chinese]
\cormark[1]
\address[1]{School of Electrical and Information Engineering, Tianjin University, Tianjin 300072, China}
\author[2]{Qi Dong}[style=chinese]
\address[2]{China Academy of Electronics and Information Technology, Beijing 100041, China}

\cortext[cor1]{Corresponding author: Guoshan Zhang (zhanggs@tju.edu.cn)}


\begin{abstract}	
This paper proposes an adaptive dynamic programming-based adaptive-gain sliding mode control (ADP-ASMC) scheme for a fixed-wing unmanned aerial vehicle (UAV) with matched and unmatched disturbances. Starting from the dynamic of fixed-wing UAV, the control-oriented model composed of attitude subsystem and airspeed subsystem is established. 
According to the different issues in two subsystems, two novel adaptive-gain generalized super-twisting (AGST) algorithms are developed to eliminate the effects of disturbances in two subsystems and  make the system trajectories tend to the designed integral sliding manifolds (ISMs) in finite time. Then, based on the expected equivalent sliding-mode dynamics, the modified adaptive dynamic programming (ADP) approach with actor-critic (AC) structure is utilized to generate the nearly optimal control laws and achieve the nearly optimal performance of the sliding-mode dynamics. Furthermore, through the Lyapunov stability theorem, the tracking errors and the weight estimation errors of two neural networks (NNs) are all uniformly ultimately bounded (UUB).
Finally, comparative simulations demonstrate the superior performance of the proposed control scheme for the fixed-wing UAV.
\end{abstract}


\maketitle

\section{Introduction}
\label{intro}

During the past decades, fixed-wing unmanned aerial vehicles (UAVs) are widely employed in military and civilian aspects, such as fire detection, disaster relief, and infrastructure inspection \cite{Wang2020Coordinated}. Compared with the rotating wing vehicles, fixed-wing UAVs have some important advantages, namely, wide flight coverage and fast flight
velocity \cite{ren2019anti}. However, the system of fixed-wing UAV also possesses some characteristics of high nonlinearity, stringent constraints, and strong coupling, which makes the flight control
extremely difficult \cite{Kang2009Linear}. Beyond that, the external disturbances during the flight process will threaten the flight stability and safety of fixed-wing UAVs.
Thus, it is a challenging task to design strongly robust, highly autonomous, and reliable flight control schemes for fixed-wing UAVs.

Recently, various control algorithms, for examples, model prediction control \cite{Dou2016Pigeon}, 
dynamic surface control
\cite{Shao2018Robust}, deep reinforcement learning method \cite{Bohn2019Deep}, and linear quadratic regulator \cite{Smith2017Disturbance}, have been employed in flight control field and promoted further study in this field.
In addition, sliding mode control (SMC)
plays an important role on designing control scheme for many complex uncertain flight systems, 
such as multiple UAVs \cite{Li2017Fault,Cao2020Distributed},  hypersonic vehicles \cite{Li2018Disturbance,Yu2017design}, and quadrotor aircraft \cite{munoz2017second}.
This algorithm ensures finite-time convergence of the sliding mode output by utilizing a discontinuous control and owns inherent insensitivity and robustness to plant uncertainties and external disturbances \cite{Ding2020Second}.
In addition, the SMC algorithms are combined with fuzzy logic \cite{Qi2020fuzzy}, 
backstepping \cite{Chen2016Robust}, and adaptive control \cite{Mofid2018Adaptive,delavari2016robust}, to improve control performance. 
Although the SMC has strong robustness, it is generally designed according to the worst case. The robustness of SMC is achieved at the cost of a high frecuency switching of the control signal, which has a negative effect in the actuators \cite{Plestan2010New}. Moreover, from the engineering viewpoint, it is a hard task to obtain the exact information of the bounds of disturbances/uncertainties. In such a situation, the larger control gains have to be set to ensure the system stability, which may lead to the overestimation of control gains and then easily cause the high frequency oscillations known as “chattering”, saturation, and higher energy consumption \cite{Lee2007Chattering}. 
The above problem motivates the development of adaptive-gain SMC algorithms. In the context of ensuring stability, adaptive-gain SMC algorithms aim at obtaining the
	control gains to be as small as possible   \cite{Li2017Adaptive,Shtessel2011Enhanced,Shtessel2010Super,Edwardsl2016Adaptive,Edwardsl2016lAdaptiveijc,Obeid2020Barrier,zhang2021fixed}. The adaptive-gain SMC algorithms can alleviate the undesired chattering effect and enhance the control performance effectively.
In the field of flight control, different gain-adaptation laws in combination with SMC algorithms are widely utilized to design flight control scheme \cite{Dong2020Adaptive,Guo2018Adaptive,Wu2019Modeling,Castaneda2017Extended,zhang2021multi}.
In the above-mentioned literatures, the gain-adaptation laws, to some extent, are able to alleviate the chattering phenomenon of SMC for systems with unknown bounded disturbances, but there also exist a few problems.
\begin{itemize} 
\item For the approaches based on the utilization of equivalent control,
the adaptive-gains in second layer are monotonically increasing and will be fixed at a constant once the states reach the predefined domain \cite{Edwardsl2016Adaptive}. The overestimation may not be avoided in the second-layer gains if dealing with the time-varying disturbances, which further affects the values of the first-layer control gains.
\item For the approaches based on increasing and decreasing the gains,  
the gains increase until
the sliding mode is achieved and then decrease until the moment it is
lost. By using this approach, the sliding variables can only converge to the neighborhood of origin, which means that the sliding mode cannot be reached any more \cite{Obeid2020Barrier,laghrouche2021barrier}. 
\item For some existing adaptive-gain SMC algorithms, the two control gains are tuned by the same gain-adaptation laws \cite{Shtessel2011Enhanced,Shtessel2010Super,Dong2020Adaptive,Guo2018Adaptive}. The parameter selection of this design is simple, but there exists a shortcoming that the control performance may be influenced when handling the state- and time-dependent disturbances simultaneously.
\end{itemize}

Under the premise of robustness and stability, the optimality is another important performance index. The objective of optimal control is to stabilize the system and meanwhile optimize a predefined cost function composed of states and control inputs \cite{Yang2020Adaptive}. 
During the past several years, many researchers have paid attentions to employing adaptive dynamic programming (ADP) approaches to address optimal control problems \cite{Wen2019Optimized,Yang2019Approximate}. Referring to \cite{Yang2018Adaptive,Zhong2016An,Mu2017Air,Tang2017Near},
ADP approaches 
 have been applied to many fields, for example, power system, single link robot arm system, flight control system and so on.
However, it is difficult to implement the ADP approach for the nonlinear systems with time-varying disturbances. In order to solve this problem, the control scheme combining integral sliding-mode control (ISMC) algorithm with ADP approach is presented \cite{Fan2016Adaptive,Zhang2019Nearly,Zhang2018Optimal,Xia2020Suboptimal}, where ADP approach is used to obtain the nearly optimal control laws for the sliding-mode dynamics.

Motivated by the above mentioned research, an adaptive dynamic programming-based adaptive-gain sliding mode control (ADP-ASMC) scheme is constructed for fixed-wing UAVs with unknown unmatched/matched disturbances. 
The main contributions of our work are summarized as follows.
\begin{enumerate} 
	\item 
	A flight control scheme constructed with SMC algorithms and ADP approach is developed for fixed-wing UAV for the first time. The proposed control scheme not only handles the effects of unknown disturbances but also optimizes the performance index when the system trajectory move on the ISMs. To some extent, the robustness and optimality are both guaranteed.  
	\item According to the different characteristics of attitude and airspeed subsystem, two novel adaptive-gain generilized super-twisting (AGST) algorithms with modified gain-adaptation laws are developed to handle the unknown bounded disturbances.
	In the designed algorithms, the prior knowledge of disturbances is not required and chattering phenomenon can be attenuated efficiently.
	\item 
	A modified ADP approach with AC structure is proposed to achieve stable and provide nearly optimal control performance of the sliding mode dynamics. The nonlinear tracking problem can be solved  effectively through segmenting an error term from the optimal performance index, and 
	no initial stabilizing control inputs are required in the training process.
\end{enumerate}

The remainder of this paper is organized as follows. The dynamic model of fixed-wing UAV and the control-oriented models (COMs) of two subsystems are given in Section \ref{section2}.  
In Section \ref{section4}, the proposed ADP-ASMC scheme
is developed. The comparative simulations and conclusion are shown in Section \ref{section5} and  Section \ref{section7}, respectively.

\section{Problem formulation}\label{section2}

\subsection{Fixed-wing UAV model}\label{section2.1}

\begin{figure}[t]
	\centering
	\includegraphics[width=7cm]{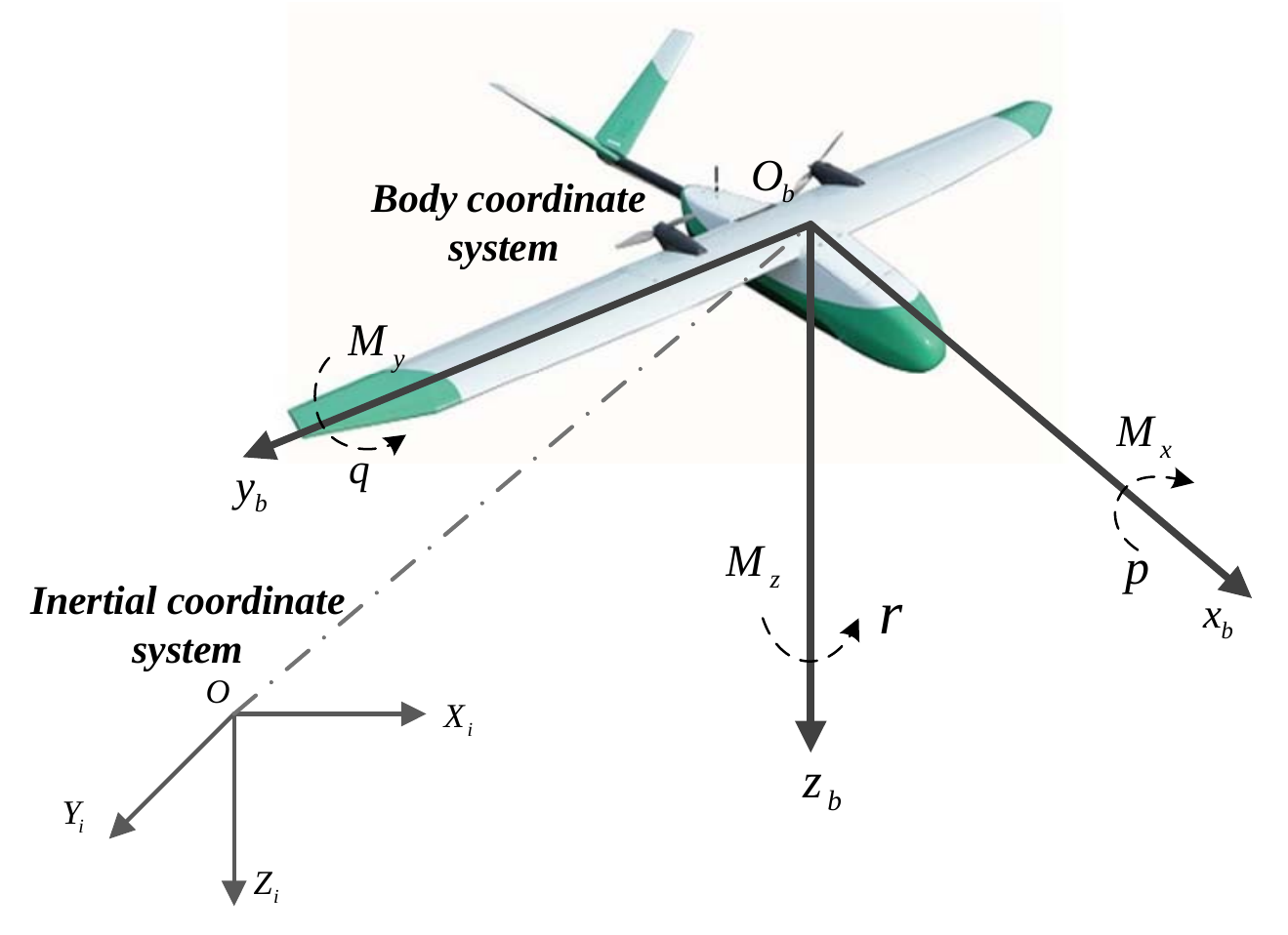}
	\caption{Referential frames configuration.}
	\label{fig-zuobiaotu}
\end{figure}

The fixed-wing UAV modelled on basis of the Newton–Euler formulation is given as \cite{Castaneda2017Extended} 
\begin{empheq}{align}
& \bm{\dot p}_n = \bm{R}_I \bm{v},\label{eq:eig1}\\
& {\bm{\dot v}} = \left( {{\bm{F}} + {\bm{T}}} \right)/m + {\bm{R}}_I{\bm{g}} - {\bm{\omega }} \times {\bm{v}}, \label{eq:eig2}\\
& {\bm{\dot \Theta }} = {{\bm{R}}_\Theta }  {\bm{\omega }}, \label{eq:eig3}\\
& {\bm{{\bm I}\dot \omega }} =  - {\bm{\omega }} \times {\bm{I\omega }} + {\bm{M}}, \label{eq:eig4}
\end{empheq}
where $\bm{p}_n=[p_x, p_y, p_z]^T$ denotes the positions of the UAV corresponding to $x_{i}$, $y_{i}$, and $z_{i}$ (shown in Fig.~\ref{fig-zuobiaotu}), $\bm{v}=[u, v, w]^T$ is the linear velocity vector, $\bm{\Theta } =[\phi,\theta,\psi]^T$ denotes the attitude angle vector, $\phi$, $\theta$, $\psi$ respectively denote roll angle, pitch angle, and yaw angle,
$\bm{\omega } =[p, q, r]^T$ stands for the angular rate vector composed of roll, pitch, and yaw angular rates, 
$\bm{T} = [T_x,0,0]^T$ represents the thrust along the body axis $x_b$, $\bm{g} = [0,0,g_z]^T$ denotes the gravity acceleration along the inertial axis $z_i$, $\bm{F} = [F_x,F_y,F_z]^T$ is the aerodynamics force vector, $\bm{M} = [M_x,M_y,M_z]^T$ is control moment vector, $M_x$, $M_y$, $M_z$ denote roll, pitch and yaw moment.
The nominal inertia matrix $\bm I \in {\mathbb{R}^{3 \times 3}}$, matrices ${\bm R_\Theta } \in {\mathbb{R}^{3 \times 3}}$ and $\bm {R_I} \in {\mathbb{R}^{3 \times 3}}$ are given as
$$\bm I{\rm{ = }}\left[ {\begin{array}{*{20}{c}}
	{I_{xx}}&{0}&{I_{xz}}\\
	{0}&{I_{yy}}&{0}\\
	{I_{xz}}&{0}&{I_{zz}}
	\end{array}} \right],$$
$${\bm R_\Theta }{\rm{ = }}\left[ {\begin{array}{*{20}{c}}
	{0}&{\cos \phi}&{-\sin \phi}\\
	{0}&{\frac{\sin \phi}{\cos \theta}}&{\frac{\cos \phi}{\cos \theta}}\\
	{1}&{\sin \phi \tan \theta}&{\cos \phi \tan \theta}
	\end{array}} \right],$$
$$\bm{R_I} = \left[ {\begin{array}{*{20}{c}}
	{\cos \theta \cos \psi }&{R_{I1}}&{R_{I2}}\\
	{\cos \theta \sin \psi }&{R_{I3}}&{R_{I4}}\\
	{ - \sin \theta }&{\cos \theta \sin \phi }&{\cos \theta \cos \phi }
	\end{array}} \right],$$
where 
\[R_{I1}=\sin \theta \cos \psi \sin \phi 
- \sin \psi \cos \phi,\] \[R_{I2}=\sin \theta\cdot \cos \psi \cos \phi 
+ \sin \psi \sin \phi,\] \[R_{I3}=\sin \theta \sin \psi \sin\phi 
+ \cos\psi \cos\phi,\]
 \[R_{I4}=\sin \theta \sin \psi \cos \phi 
- \cos \psi \sin \phi.\]

\subsection{Control-oriented model}\label{section2.2}
The control-oriented models (COMs) of fixed-wing UAV are established in this subsection.
The attitude controller is designed according to the attitude subsystem \eqref{eq:eig3} and \eqref{eq:eig4}.
In our work, the attitude angle $\bm \Theta$ and angular rate $\bm \omega$ are both measurable. 
Considering the influence of unknown matched disturbances $\bm{\Delta d_m}(t)$ and unmatched disturbances $\bm{\Delta d_u}(t)$,
\eqref{eq:eig3} and \eqref{eq:eig4} are rewritten as 
\begin{align}\label{eq5}
{\bm{\dot \Theta }} =& {{\bm{R}}_\Theta } {\bm{\omega }}+\bm{\Delta d_u(t)}\nonumber\\
{\bm{\dot \omega }} =&  - {{\bm{{\bm I}}}^{ - 1}}{\bm{\omega }} \times {\bm{I\omega }} + {{\bm{{\bm I}}}^{ - 1}}({\bm{M}} + {\bm{\Delta d_m}(t)}).
\end{align}
To ease the notation, we denote $\bm {\Delta d_u}(t)=\bm {\Delta d_u}$ and $\bm {\Delta d_m}(t)=\bm {\Delta d_m}$ in the sequel.	
The reference attitude command $\bm{\Theta _d} \in {\mathbb{R}^{3 \times 1}}$ and its derivative $\bm{\dot \Theta }_{\bm{d}}$ are smooth, bounded, and known. Define that the attitude tracking error is ${{\bm{e}}_{\bm{\Theta }}}{\bm{ = \Theta  - }}{{\bm{\Theta }}_{\bm{d}}} = [e_{\phi},\\e_{\theta},e_{\psi}]^T$ and the dynamic of $\bm {e_\Theta}$ is 
\begin{equation*}
{{\bm{\dot e}}_{\bm{\Theta }}} = {{\bm{R}}_\Theta } {\bm{\omega }} - {{\bm{\dot \Theta }}_{\bm{d}}} + {\bm{\Delta }}{{\bm{d}}_u}.
\end{equation*}
Let ${{\bm{\dot e}}_{\bm{\Theta }}} = {{\bm{z}}_{\bm{\Theta }}}$ and we can obtain
\begin{equation}\label{eq7}
\begin{array}{l}
{{{\bm{\dot e}}}_{\bm{\Theta }}}{\bm{ = }}{{\bm{z}}_{\bm{\Theta }}}\\
{{{\bm{\dot z}}}_{\bm{\Theta }}} = {\bm{G}}\left( \bm {z_\Theta} \right) + {{\bm{R}}_\Theta }{{\bm{{\bm I}}}^{ - 1}}{\bm{M}}- {{{\bm{\ddot \Theta }}}_{\bm{d}}} + {{\bm{\Delta }}_{\bm{a}}}+ {\bm{\Delta }}{{{\bm{\dot d}}}_u}
\end{array}
\end{equation}
where ${\bm{G}}\left( {{{\bm{z}}_{\bm{\Theta }}}} \right) =\ {{{\bm{\dot R}}}_\Theta } \bm \omega - {{\bm{R}}_\Theta }{{\bm{{\bm I}}}^{ - 1}}{\bm \omega}\times\bm{I \omega}$ and ${{\bm{\Delta }}_{\bm{a}}} =    {{\bm{R}}_\Theta }{{\bm{{\bm I}}}^{ - 1}}\bm\Delta {\bm d_m}$ is the lumped disturbance vector related to the states. 
\eqref{eq7} is the COM of attitude subsystem, which is a second-order multiple-input multiple-output (MIMO) system.
 
\begin{remark}\label{r1}
	The pitch angle $\theta$ cannot be equal to $ \pm \frac{\pi}{2}$ such that the matrix $\bm{R_\Theta}$ is always invertible and the control moment vector $\bm M$ is nonsingular. 
\end{remark}

\begin{assumption}
	Assume that the matched disturbance $\bm {\Delta}\bm {d_m}$ is bounded by an unknown positive constant $\delta_{dm}$, i.e., $\left\|\bm {\Delta}\bm {d_m}\right\|\\\le \delta_{dm}$. 
	The unmatched disturbance ${\bm{\Delta }}{{\bm{d}}_u}$ is bounded by an unknown positive constant $\delta_{u0}$, namely, $\bm \left\|{\bm{\Delta }}{{\bm{d}}_u}\right\|\le\delta_{u0}$.
	Beyond that, ${\bm{\Delta }}{{\bm{d}}_u}$ is also bounded by unknown Lipschitz constants $\delta_{du1}$ and $\delta_{du2}$, namely, $\left\| {{\bm{\Delta }}{{{\ddot{\bm d}}}_u}} \right\|\le\delta_{du1}$ and $\left\| {{\bm{\Delta }}{{{\bm{\dddot d}}}_u}} \right\|\le\delta_{du2}$.
	\label{assum2}
\end{assumption}

In fixed-wing UAV, the linear velocity vector $\bm v $ in \eqref{eq:eig2} is controlled through the airspeed controller which is designed according to the airspeed subsystem \cite{Castaneda2017Extended,Stevens2015Aircraft}. Following \eqref{eq:eig2}, the dynamic of airspeed subsystem is derived as
\begin{equation}\label{eq8}
\dot V = \frac{{{T_x}\cos \alpha\cos \beta  - D}}{m} - {g_v} + {\Delta _V}(t),
\end{equation}
where $V$ denotes airspeed, $\alpha  = \arctan \left( {\frac{w}{u}} \right)$ and $\beta  = \arcsin \left( {\frac{u}{v}} \right)$ respectively represent angle of attack and sideslip angle, $D$ represents drag, ${\Delta _V}$ is external disturbance of airspeed subsystem, $g_v$ is the term that is relevant to attitude subsystem and is given as
\begin{equation*}
\begin{split}
{g_v} = & \ g( - \cos \alpha \cos \beta \sin \theta  + \sin \beta \sin \phi \cos \theta  \\
&+ \sin \alpha \cos \beta \cos \phi \cos \theta  ).
\end{split}
\end{equation*}

\begin{assumption}
	The external disturbance ${\Delta _V}$ is unknown but bounded, i.e., $\left| {{{ \Delta }_V}} \right| \le {\delta _{V}}$. Besides, the derivative $\dot{\Delta }_V$ and second derivative $\ddot{\Delta }_V$ are also bounded, i.e., $\left| {{{\dot \Delta }_V}} \right| \le {\delta _{V0}}$ and $\left| {{{\ddot \Delta }_V}} \right| \le {\delta _{V1}}$. ${\delta _{V}} $, ${\delta _{V0}} $, and ${\delta _{V1}} $ are unknown positive constants.
	\label{assum3}
\end{assumption}

$V_d$ is the reference airspeed command and the airspeed tracking error is 
$e_V=V-V_d$.
The dynamic of $e_V$ is
\begin{equation}\label{eq9}
{\dot e_V} = \frac{{{T_x}\cos \alpha \cos \beta  - D}}{m} - {g_v} - {\dot V_d} + {\Delta _V}\left( t \right),
\end{equation}
which is a first-order single-input single-output (SISO) system.
It should be noted that $\alpha$ and $\beta$ cannot be equal to $\pm\frac{\pi}{2}$ to avoid the singularity of $T_x$.

Following the above analysis, \eqref{eq7} and \eqref{eq9} are the COMs utilized to design control scheme.

\textit{Control Objective:} The aim of our work is to design the control laws $\bm M$ and $T_x$ for fixed-wing UAV such that the attitude angle $\bm \Theta$ and airspeed $V$ can track the reference command $\bm {\Theta_d}$ and $V_d$ under the effects of $\bm {\Delta d_u}$, $\bm {\Delta d_m}$, and $\Delta_V$.

\subsection{Preliminaries}
\label{section2.3}

Before the detailed control scheme design, Definition \ref{defi1},  Notation \ref{notation1}, and Proposition \ref{le1} are presented in this subsection.
\begin{definition}\textit{\cite{Zhang2018Optimal}}\label{defi1}
	The equilibrium point $\bar x_0$ for system
	\begin{equation}\label{lemma}
	\dot x\left( t \right) = f\left( {x\left( t \right)} \right), x\left( 0 \right) = {x_0}
	\end{equation}
	is said to be uniformly ultimately bounded (UUB) if there exists a compact set $\Omega  \subset {\mathbb{R}^n}$ so that for all ${\bar x_0} \in \Omega $ there exists a bound $\rm B$ and a time $T\left( {{\rm B}, {{\bar x}_0}} \right)$ such that $\left\| {x\left( t \right) - {{\bar x}_0}} \right\| \le {\rm B}$ for all $t \ge {t_0} + T$.
\end{definition}
\begin{notation}\label{notation1}
	$\mathbb{R}$, $\mathbb{R}^n$, and $\mathbb{R}^{m\times n}$ denote the set of real numbers, the Euclidean space of real n-vectors, and 
	the space of $m \times n$ real matrices,respectively.
	For a given state vector $\bm x =[x_1,x_2,...,x_n]^T$, define the multivariable sign function $\textbf{sign}\left( \bm x \right) = \frac{ \bm x}{{\| { \bm x} \|}}$ and $\left\| {\textbf{sign}\left( {\bm x} \right)} \right\| = 1$. $\lceil {\bm  x} \rfloor^{\rho}=\| { \bm x} \|^{\rho}\textbf{sign}(\bm  x)$. $\| \bm x \| = \sqrt {{\bm x^T}\bm x} $ denotes the Euclidean norm of vectors. $\lceil {\bm x} \rfloor^{\rho}$ is continuous for any $\rho>0$ at zero and its value is understood in the sense of Filippov \cite{Filippov2013Differential}. For a given state $x$, $\lceil { x} \rfloor^{\rho}=\left| {x} \right|^\rho\sign(x)$, where $\left| \cdot \right|$ represents the absolute value of a scalar variable and $\sign( \cdot )$ is the symbolic function.
	For a given matrix $\bm A$, define that ${\lambda _{\min}}\left( {{\bm A}} \right)$ and ${\lambda _{\max}}\left( {{\bm A}} \right)$ are the minimum and maximum eigenvalues of ${\bm A}$. $\otimes$ represents the Kronecker product between two matrices.
\end{notation}

\begin{proposition}\label{le1}
For a symmetric positive definite matrix $ \bm A\in \mathbb{R}^{m\times m}$ and $n$-dimension identity
	matrix $\bm I_n$, the matrix $\tilde{ \bm A}=\bm A\otimes \bm I_n\in \mathbb{R}^{mn\times mn}$ is also a symmetric positive
	definite matrix.
\end{proposition}

\section{Main Results}\label{section4}

In this section, the detailed design process of the proposed ADP-ASMC scheme is introduced.

\subsection{Sliding mode control design}\label{section3.2}

For attitude subsystem \eqref{eq5} and airspeed subsystem \eqref{eq8}, there exist the unknown disturbance $\bm {\Delta d}_m$, $\bm {\Delta d}_u$, and $ \Delta_V$, which are relevant to time $t$. 
It is difficult for the existing ADP approaches to solve the time-varying optimal control problem directly \cite{Fan2016Adaptive}.
Thus, according to the different issues in two subsystems, two novel adaptive-gain generalized super-twisting (AGST) algorithms are combined with integral sliding manifolds (ISMs) to eliminate the effects of $\bm {\Delta d}_m$, $\bm {\Delta d}_u$, and $ \Delta_V$. Then, a modified ADP approach is developed in Subsection \ref{sec.c} to generate the nearly optimal control laws for the sliding mode dynamics. In this subsection, the sliding mode control laws for two subsystems are designed.

\subsubsection{ISM-based AMGST control for attitude subsystem}\label{sec1)}
The control moment $\bm M$ in attitude subsystem consists of two parts, that is,
\begin{equation}\label{equation21}
\bm M=\bm M_s+\bm M_a,
\end{equation}
where $\bm M_s$ is designed in basis of ISM-based AMGST algorithm, $\bm M_a$ is generated via ADP approach.

In this part, the design process of $\bm M_s$ is introduced. First, define an integral sliding manifold as
\begin{equation}\label{eq22}
\bm S =\bm {z_\Theta} -\int_{0}^{t}{ \bm {R_\Theta I}^{-1}\bm M_a- {{{\bm{\ddot \Theta }}}_{\bm{d}}}}d\tau_t. 
\end{equation}
The derivative of $\bm S$ is
\begin{equation}\label{eq23}
\dot{\bm S} =\bm{G({\bm z}_\Theta)}+\bm {R_\Theta I}^{-1}\bm M_s+\bm{ \Delta_a}+\bm \Delta \dot{\bm d}_u. 
\end{equation}
According to \eqref{eq23}, $\bm M_s$ is designed as
\begin{align}\label{sliding}
\bm M_s =&\bm I \bm {R_\Theta}^{-1}\big(-k_{1} \bm \Phi_1 ({\bm S})+\bm z_1-\bm{G({\bm z}_\Theta)}\big)\nonumber\\
\dot{\bm z}_1=&-k_{20}L\bm \Phi_2 ({\bm S}),
\end{align}
where 
$\bm \Phi_1 ({\bm S})=\lceil {{\bm{S}}} \rfloor^{1/2}+{\bm{S}}$ and $\bm \Phi_2 ({\bm S})={\bm \Phi_1'}\bm{\Phi}_1=\frac{1}{2}\lceil {\bm{S}} \rfloor^{0}+\frac{3}{2}\lceil {\bm{S}} \rfloor^{1/2}+{\bm{S }}$, 
$k_{1}$ and $L$ are adaptive gains, which are adapted as
\begin{equation}\label{jeq24}
{\dot k_1} = \left\{ {\begin{array}{*{20}{c}}
	{{\kappa_1 \left\|{ {\bm {S }}} \right\|}+\kappa_0,\ if \left\|{ {\bm {S }}} \right\| \ne 0}\\
	{0, otherwise}
	\end{array}} \right.	
\end{equation}
\begin{equation}\label{eq24}
\left\{ {\begin{array}{*{20}{c}}
	{\begin{array}{*{20}{c}}
		{\begin{array}{*{20}{c}}
			{{L} = {L_{0}} + \Delta {L}}\\
			{{\Delta {\dot L}} = - \lambda \left( t \right)\sign\left( {{e_\Delta }} \right)}
			\end{array}}\\
		{{e_\Delta } = 0.5{L} - \frac{1}{{a l}}\left\| {{{\bar u}_{eq}}} \right\| - \varepsilon }
		\end{array}}\\
	{\lambda \left( t \right) = {\lambda _0} + r\left( t \right)}
	\end{array}} \right.
\end{equation}
\begin{equation}\label{eq25}
\dot r\left( t \right)  = \left\{ {\begin{array}{*{20}{c}}
	{{\bar r \left| {{e_\Delta }} \right|\sign\left( {\left| {{e_\Delta }} \right| - \bar e} \right),{\kern 1pt} {\kern 1pt} {\kern 1pt} r\left( t \right) > {r_m}}}\\
	{{{r_m},{\kern 1pt} {\kern 1pt} otherwise}}
	\end{array}} \right.
\end{equation}
where $k_{20}$, $\kappa_{1}$, $\kappa_{0}\in (0,1)$, $\bar e\in(0,1)$, ${L_{0}}$, $\bar r $, ${r_m}$, $\lambda _0$ are positive constants and $0 < a < 1/ l < 1$, $\varepsilon $ is a small positive constant.
${\bar u_{eq}}$, which can be obtained by low-pass filtering the term $\frac{{{k_{20}}L}}{2} \lceil {{\bm S}} \rfloor^{0}$, represents the approximate value of the equivalent control ${u_{eq}}$. It should be noted that the equivalent control $u_{eq}$ is just theoretical control. The error between $u_{eq}$ and its approximation $\bar u_{eq}$ can be very small via selecting an appropriate time constant $\tau $ \cite{Edwardsl2016Adaptive,Utkin2013Adaptive}.
\begin{assumption}\label{ass-a2}	
	For lumped disturbance $\bm {\Delta_a}$, because it is related to the state $\bm \Theta$, and $\bm \Phi_1({\bm S})$ is a function associated with $\bm \Theta$ (since ${\bm S}$ is related to $\bm \Theta$), 
	it is reasonable to assume that there exists an unknown positive constant $\delta_a$ to make $\bm {\Delta_a}$ satisfy $\left\|\bm {\Delta_a}\right\|\le \delta_a\left\|\bm \Phi_1({\bm S})\right\|$.
\end{assumption}

Substituting \eqref{sliding} into \eqref{eq23} yields
\begin{align}\label{eqS}
\dot{ {\bm S}}=&-k_1\bm{\Phi}_1({\bm S})+\bm z+\bm \Delta_a\nonumber\\
\dot{\bm z}=&-k_{20}L\bm{\Phi}_2({\bm S})+{\bm \Delta}\ddot{\bm d}_u
\end{align}
where 
$\bm z=\bm z_1+{\bm \Delta}\dot{\bm d}_u$. 
\begin{remark}\label{remark_aa}
	Since the attitude subsystem is a second-order MIMO system, we employ the multivariable generalized super-twisting algorithm. The multivariable design can avoid the necessity for the decoupled design with three SISO structures and it also exploits the coupling and inherent functional redundancy in fixed-wing UAV, which improves the safety of flight control.
\end{remark}

Compared with standard super-twisting algorithm (STA), the generalized super-twisting algorithm (GSTA) can offer stronger robustness since the linear growth term in $\bm \Phi_1$ helps to counteract
the state-dependent disturbance \cite{Castillo2018Super}.
Furthermore, the two gains $k_1$ and $k_{20}L$ in $\bm M_s$ are tuned via different gain-adaptation laws, respectively.
The reason is that the disturbance $\bm \Delta_a$ and $\bm \Delta {\bm d}_u$ are different, namely, $\bm \Delta_a$ is a state-dependent disturbance and $\bm \Delta {\bm d}_u$ is just a time-dependent disturbance. The gain-adaptation law of $k_1$ can counteract the state-dependent disturbance $\bm {\Delta_a}$. Under the adjustment of gain-adaptation law \eqref{eq24}-\eqref{eq25}, $L$ changes with time-dependent disturbance $\bm \Delta{\ddot{\bm d}}_u$ in real-time. Besides, compared to the adaptive-gain super-twisting (AST) in \cite {Shtessel2011Enhanced}, the proposed AMGST guarantees that the sliding mode manifold can converge to origin in finite time. However, the sliding variables under the AST can only converge to the finite domain of origin in theory.

\begin{remark}\label{two adaptive}
The gain-adaptation law of $r(t)$ is different from that in \cite{Edwardsl2016Adaptive}. 
		The overestimation of $r(t)$ is considered in our work. Once the sliding mode associated with $e_\Delta$ is achieved, $r(t)$ shall start reducing and keep at a small level. Due to the decreasing of $r(t)$, the first-layer adaptive gain $L$ will be smaller than that in \cite{Edwardsl2016Adaptive}. 
\end{remark}

\begin{theorem}\label{theorem2}
	Consider system \eqref{eq7} with Assumptions \ref{assum2}- \ref{ass-a2}. Let the integral sliding manifold $\bm S$, sliding mode control law $\bm M_s$ and gain-adaptation law be designed as \eqref{eq22}, \eqref{sliding}, and \eqref{jeq24}-\eqref{eq25}, respectively. Then, the system \eqref{eq7} can reach the integral sliding manifold in finite time.
\end{theorem}

\begin{pf}	

For system \eqref{eqS}, suppose that a gain-adaptation law has already been devised for updating $L$ which is differentiable, bounded and satisfies $L>\max\left\{ {L_0, 2\left\| {\bm{\Delta} \ddot {\bm d}_u} \right\|} \right\}$. Under the above premise,  
we have $L>2\delta_{du1}$. Define $\eta=\frac{2\delta_{du1}}{L}\in(0,1)$ and $\bm X =\big [\bm X_1,\bm X_2\big]^T=\big [\bm \Phi_1,\bm z\big]^T\in \mathbb{R}^6$. The derivative of $\bm{X}$ is
\begin{equation}\label{E2}
\left[ {\begin{array}{*{20}{c}}
	{{{\dot {\bm X}}_1}}\\
	{{{\dot {\bm X}}_2}}
	\end{array}} \right]=- {p_e}\bm A_x \otimes \bm {I_3}\left[ {\begin{array}{*{20}{c}}
	{{\bm X_1}}\\
	{{\bm X_2}}
	\end{array}} \right],
\end{equation}
where $\bm A_x=\left[ {\begin{array}{*{20}{c}}
	{{k_1} - \frac{{{\bm{\Delta_a}^T}\textbf{sign}\left( {{\bm \Phi_1}} \right)}}{{\left\| {{\bm \Phi_1}} \right\|}}}&-1\\
	{{k_{20}L} - \frac{{ {\bm{\Delta} \ddot {\bm d}^T_u}\textbf{sign}\left( {{\bm \Phi_2}} \right)}}{{\left\| {{\bm \Phi_2}} \right\|}}}&0
	\end{array}} \right]$, $p_e=\bm X_1'=\bm \Phi_1'=\frac{1}{{2\left\| {\bm {S}}   \right\|^{1/2}}} + 1$, and $\left| \frac{1}{{p_e}} \right| \le 1$.

To examine the stability of \eqref{E2}, a Lyapunov candidate is selected as 
\begin{equation}\label{V_outer}
V_X=V_{X0}+\frac{(k_1-\bar {k}_1)^2}{2\Gamma},
\end{equation}
where $V_{X0}=\frac{1}{2}\bm X^T \bm P \bm X$, $\Gamma>0$, and ${\bm P} =
\left[ {\begin{array}{*{20}{c}}
	{{p_1}}&{ - 1}\\
	{ - 1}&{{p_2}}
	\end{array}} \right]\otimes \bm I_3=\bar {\bm P}\otimes \bm I_3$ with $p_1>0$, $p_1p_2>1$. According to Proposition \ref{le1},  ${\bm P}$ is positive definite. $\bar {k}_1$ is the upper bound of $k_1$ (the  boundedness of $k_1$ will be analyzed later). The proof of Lyapunov candidate $V_X$ is divided into two steps. 

\textit{{Step 1:}} The stability of $V_{X0}$ is analyzed. Taking derivative of $V_{X0}$, we have
\begin{equation}\label{dV_outer}
\dot{V}_{X0}=-{ p_e}\bm X^T\bm Q \bm X,
\end{equation}
where $\bm Q=\bm A^T_x\bm P+\bm P \bm{A}_x=\left[ {\begin{array}{*{20}{c}}
	{q_1}&{q_2}\\
	{q_2}&{q_3}
	\end{array}} \right]\otimes \bm{I_3}$ and 
$$q_1=\tilde{k}_1p_1-\tilde{k}_2,$$ $$q_2=\frac{1}{2}(-p_1-\tilde{k}_1+p_2\tilde{k}_2),$$ $$q_3=1 $$
with $\tilde{k}_1={k_1} -\frac{{{\bm{\Delta_a}^T}\textbf{sign}\left( {{\bm \Phi_1}} \right)}}{{\left\| {{\bm \Phi_1}} \right\|}}$ and  $\tilde{k}_2={k_{20}L} - \frac{{ {\bm{\Delta} \ddot {\bm d}^T_u}\textbf{sign}\left( {{\bm \Phi_2}} \right)}}{{\left\| {{\bm \Phi_2}} \right\|}}$. Noting the fact that $\left| \frac{1}{{p_e}} \right| \le 1$ and $\left\| {\frac{{1}}{{\bm{\Phi}_2}}} \right\|\le2$, the value ranges of $\tilde{k}_1$ and $\tilde{k}_2$ are deduced, that is, 
$${{\tilde k}_1} \in \left[\tilde k_{1m}, \tilde k_{1M}\right]=\left[ {k_1-\delta_a, k_1+\delta_a} \right],$$
$${{\tilde k}_2} \in \left[\tilde k_{2m}, \tilde k_{2M}\right]=\left[ {k_{20}L-2\delta_{du1}, k_{20}L+2\delta_{du1}} \right].$$
To make $\dot V_{E0}$ negative definite, $\bm Q$ should satisfy
\begin{empheq}{align}
 {k}_1>&\frac{\frac{k_{20}}{\eta}+1}{\frac{p_1}{2\delta_{du1}}}+\delta_a,\label{eq:eig21}\\
 -\frac{1}{4}p_1^2&+\frac{1}{2}(\tilde k_1+p_2\tilde k_2)p_1-\tilde k_2-\frac{1}{4}(\tilde k_1-p_2\tilde k_2)^2>0. \label{eq:eig22}
\end{empheq}
\eqref{eq:eig22} can be regard as a quadratic equation with respect to $p_1$. To make $\bm Q$ positive definite , the discriminant of \eqref{eq:eig22}, i.e., $\Delta_Q=\tilde k_2(p_2\tilde k_1-1)$, has to be greater than zero.
Thus, we have
\begin{empheq}{align}
&k_{20}>\eta,\label{25}\\
&k_{1}>\frac{1}{p_2}+\delta_a,\label{26}
\end{empheq}
and the solution of inequality \eqref{eq:eig22} is
$$p_1\in\left(p_-, p_+\right)
=\left(\tilde k_{1}+p_2\tilde k_{2}-2\sqrt {\Delta_{Q}}, \tilde k_{1}+p_2\tilde k_{2}+2\sqrt {\Delta_{Q}}\right).$$
To ensure $p_1\in\left(p_-, p_+\right)$ is a non-empty intersection set, the following condition has to be satisfied.
\begin{align}\label{27}
&\tilde k_{1M}+p_2\tilde k_{2M}-2\sqrt {\tilde k_{2m}(p_2\tilde k_{1m}-1)}<\nonumber\\
&\tilde k_{1m}+p_2\tilde k_{2m}+2\sqrt {\tilde k_{2m}(p_2\tilde k_{1m}-1)}.
\end{align} 
Through several algebraic manipulations of \eqref{27}, we deduce
\begin{equation}\label{28}
k_1>\frac{k_{\varepsilon}+1}{p_2}+\delta_a
\end{equation}
with $k_{\varepsilon}=\frac{\frac{\delta_a^2}{2\delta_{du1}}+2\delta_{a}p_2+2p_2^2\delta_{du1}}{4(\frac{k_{20}}{\eta}-1)}$.
When $k_1$ satisfies 
\begin{equation}\label{29}
k_1>\max\left[ {\frac{\frac{k_{20}}{\eta}+1}{\frac{p_1}{2\delta_{du1}}}, \frac{k_{\varepsilon}+1}{p_2}} \right]+\delta_{a},
\end{equation} 
$\dot{V}_{X0}$ is negative definite. 

Following the above analysis, we deduce that $\bm Q$ is positive definite when $k_{20}$ and $k_{1}$ satisfy \eqref{25} and \eqref{29} and have 
$$\lambda_{\min}(\bm Q)\left\| {\bm X} \right\|^2\le\bm X^T \bm Q \bm X \le\lambda_{\max}(\bm Q)\left\| {\bm X} \right\|^2.$$
Noting the fact that $\left\| {\bm X_1} \right\|\le\left\| {\bm X} \right\|$ and $\lambda_{\min}(\bm P)\left\| {\bm X} \right\|^2\le\bm X^T \bm P \bm X \le\lambda_{\max}(\bm P)\left\| {\bm X} \right\|^2$, \eqref{dV_outer}
can be transformed into
\begin{align}\label{31}
\dot{V}_{X0}
\le&-\nu_1 V_{X0}^{1/2}-\nu_2 V_{X0}
\end{align}
with $\nu_1=\frac{\lambda_{\min}(\bm Q)\sqrt{\lambda_{\min}(\bm P)}}{2\lambda_{\max}(\bm P)}$
and $\nu_2=\frac{\lambda_{\min}(\bm Q)}{\lambda_{\max}(\bm P)}$.
Following the Theorem 4.2 in \cite{Bhat2000Finite} and the assumption of $L$, if $k_1$ and $k_{20}$ are selected to satisfy \eqref{29} and \eqref{25}, the system \eqref{E2} is finite-time stable.

From the result in \eqref{31}, the stability of $V_X$ is analyzed. Under the condition that the initial value of $k_1$ is greater than $k_m$, the derivative of \eqref{V_outer} is
\begin{align}\label{32}
\dot{V}_X
\le&-\nu_1 V_{X0}^{1/2}-\nu_2 V_{X0}+\frac{1}{\Gamma}(k_1-\bar {k}_1)(\kappa_1\left\| {\bm {{\bm S} }} \right\|+\kappa_0)\nonumber\\
&+\beta_K\left| {k_1-\bar{k}_1} \right|-\beta_K\left| {k_1-\bar{k}_1}\right|+\frac{\bar{k}_1}{\Gamma}\kappa_1\left\|{\bm S}\right\|\nonumber \\
&+\beta_M\left| {k_1-\bar{k}_1} \right|\bar{k}_1-\beta_M ({k_1-\bar{k}_1})^2-\frac{k_1}{\Gamma}\kappa_1\left\|{\bm S}\right\|\nonumber \\
\le&-\nu_3{V}_X^{1/2}-\nu_4{V}_X-\varsigma\left| {k_1-\bar{k}_1} \right|,
\end{align}
where $\beta_K>0$, $\beta_M>0$, $\nu_3=\min\left\{ {\nu_1,\beta_K} \right\}$, $\nu_4=\min\left\{ {\nu_2,\beta_M} \right\}$, and 
$$\varsigma=-\beta_K+\frac{\kappa_0}{\Gamma}-\beta_M\bar k_1.$$ 
There always exists $\Gamma=\frac{\kappa_0}{\beta_k+\beta_M\bar k_1}$, which yields $\varsigma=0$. Then, inequality \eqref{32} is rewritten as
\begin{equation}\label{46}
\dot{V}_X\le-\nu_3{V}_X^{1/2}-\nu_4{V}_X,
\end{equation}
which means that $ {\bm S}$ can converge to the origin in finite time. 

It is worth noting that for the finite time convergence, $k_1$ and $k_{20}$ must satisfy inequality \eqref{29} and \eqref{25} under the assumption $L>\max\left\{ {L_0, 2\left\| {\bm{\Delta} \ddot {\bm d}_u} \right\|} \right\}$. Namely, when $\left\|{\bm{S}} \right\| \ne 0$, $k_1$ will increase at the rate $(\kappa_1\|\bm S\|+\kappa_0)$ until the condition \eqref{29} is satisfied. Then, $\left\| {\bm{S}} \right\|$ can converge in finite time.

According to the above analysis, it can be concluded that $k_1$ is not monotonically increasing all the time. When $\left\|{\bm S}\right\|$ converges to the origin, $\dot{k}_1$ is equal to $0$ and then ${k_1}$ will remain unchanged. Therefore, ${k_1}$ is bounded by a positive constant $\bar k_1$. The boundedness of ${k_1}$ is ensured.  

\textit{Step 2:} In this step, the stability of adaptive gain $L$ is analyzed. On the reaching phase, $L<2\delta_{du1}$ and then we have $2\| {\bar u_{eq}} \|>L$. Thus, there exists
\begin{equation}\label{36}
{{e_\Delta } = 0.5{L} - \frac{1}{{a l}}\left\| {{{\bar u}_{eq}}} \right\| - \varepsilon }<0\nonumber.
\end{equation}
Obviously, ${\Delta}\dot{L}>0$ and $L$ will increase at the rate $\lambda(t)$. After $T\le t_0+\frac{2\delta_{du1}-L_0}{\lambda_0}$, $L$ will be greater than $2\delta_{du1}$. 

Now consider the following Lyapunov function.
\begin{equation}\label{37}
V_L=\frac{1}{2}e_\Delta^2+\frac{1}{2\Gamma_l}(r-r^*)^2,
\end{equation}
where $r^*$ represents the upper bound of $r$ and $\Gamma_l$ is an appropriate positive constant. In view of \eqref{eq24} and \eqref{eq25}, it is easy to deduce that
\begin{align}\label{38}
\dot V_L
\le&-\nu_l\left| {e_\Delta} \right|-\beta_l\left| {r-r^*} \right|+\beta_l\left| {r-r^*} \right|\nonumber\\
&-\frac{1}{\Gamma_l}\left| {r-r^*} \right|\bar r \left| {{e_\Delta }} \right|\sign\left( {\left| {{e_\Delta }} \right| - \bar e} \right)
\end{align}
with $\nu_1=0.5(\lambda_0+r)-\frac{1}{al}\delta_{du2}$ and $\beta_l>0$. 
There exists a condition that $\lambda_0+r>\frac{2}{al}\delta_{du2}$ to make sure $\nu_l>0$.
Define $\varsigma_l=-\big(\beta_l-\frac{1}{\Gamma_l}\bar r \left| {{e_\Delta }} \right|\sign\left( {\left| {{e_\Delta }} \right| - \bar e} \right)\big)$ and we deduce
\begin{align}\label{39}
\dot V_L\le&-\bar \nu_lV_L^{1/2}-\varsigma_l\left| {r-r^*} \right|.
\end{align}  
where $\bar \nu_l=\min\left\{ {\nu_l, \beta_l} \right\}$. When $\left| {{e_\Delta }} \right|>\bar e$, $r$ will increase. If $\Gamma_l<\frac{\bar r\bar e}{\beta_l}$, $\varsigma_l$ will be positive and then the inequality $\dot V_L\le-\bar \nu_lV_L^{1/2}$ holds. According to the Theorem 4.2 in \cite{Bhat2000Finite}, $e_\Delta$ can converge to the prescribed small interval $\left| {{e_\Delta }} \right|<\bar e$ in finite time. 
When $\left| {{e_\Delta }} \right|<\bar e$, $\varsigma_l$ is negative, which results in the sign indefinite of $\dot V_L$. $r$ will decrease and $e_\Delta$ may increase. The process will go back to the case that  $\left| {{e_\Delta }} \right|>\bar e$. Accordingly, $e_\Delta$ is restricted to a finite domain of origin, namely, $\left|e_\Delta\right|< \bar e_1$ with $\bar e_1>\bar e $.

The proof of Theorem \ref{theorem2} is completed.
\end{pf}

Following Theorem \ref{theorem2}, the state trajectory of system \eqref{eq7} reaches the sliding manifold $\bm S$ under the control law \eqref{sliding}. 
According to \eqref{eq23} and $\dot{\bm S}=0$, the equivalent control law is
$$\bm M_{seq}=-\bm I\bm {R_\Theta}^{-1}(\bm \Delta \dot{\bm d}_u+\bm {\Delta_a}+\bm G(\bm {z_\Theta})).$$
Substituting $\bm M_{seq}$ into \eqref{eq7}, we derive
\begin{equation}\label{peq26}
\begin{array}{l}
{{{\bm{\dot e}}}_{\bm{\Theta }}}{\bm{ = }}{{\bm{z}}_{\bm{\Theta }}}\\
{{{\bm{\dot z}}}_{\bm{\Theta }}} =   {{\bm{R}}_\Theta }{{\bm{{\bm I}}}^{ - 1}}{\bm{M}_a}- {{{\bm{\ddot \Theta }}}_{\bm{d}}}. 
\end{array}
\end{equation}
Define $\bm E=[\bm{e_\Theta}^T, \bm{z_\Theta}^T]^T\in\mathbb{R}^6$, $\bm F(\bm{z_\Theta})=[\bm{z_\Theta}^T,\bm{0}_{1\times3}]^T\in \mathbb{R}^6$, $\bar{\bm \Theta}_d=[\bm{0}_{1\times3}, \ddot{\bm \Theta}^T_{\bm d}]\in\mathbb{R}^6$, and $\bar{\bm g}=[\bm {0}_{3\times3}, (\bm{R_\Theta I}^{-1})^T]^T\\\in\mathbb{R}^{6\times3}$. \eqref{peq26} is rewritten as
\begin{equation}\label{peq27}
\dot{\bm E}=\bm F(\bm{z_\Theta})+\bar{\bm g}\bm M_a-\bar{\bm \Theta}_d.
\end{equation}
According to \eqref{peq26} and \eqref{peq27}, $\bm {\Delta_a}$ and $\bm {\Delta d}_u$ are completely compensated by $\bm M_s$. Without the effect of time-varying disturbances, the optimal control problem for system \eqref{peq27} can be regard as time-invariant, and then the ADP approach is used to design $\bm M_a$, whose design process is shown in Subsection \ref{sec.c}.

\subsubsection{ISM-based AGST control for airspeed subsystem}\label{sec2}
 
Similar to the design of $\bm M$ in attitude subsystem, the thrust $T_x$ is also composed of two parts, namely,
\begin{equation}\label{thrust1}
T_x =T_{xs}+T_{xa},
\end{equation}
where $T_{xs}$ is generated via the ISM-based AGST algorithm. $T_{xa}$ is generated via ADP approach. The design process of $T_{xs}$ is detailed in this part.
First, the ISM of airspeed subsystem is designed as
\begin{equation}\label{tx}
S_V=e_V(t)-\int_{0}^{t}{-g_v-\dot {V}_d+\frac{{\cos \alpha \cos \beta{T_{xa}}-D }}{m}}d\tau_t.
\end{equation}
Taking the derivative of $S_V$, we have
\begin{equation}\label{eq32}
\dot{S}_V=\frac{{\cos \alpha \cos \beta{T_{xs}} }}{m}+\Delta_V
\end{equation} 
and $T_{xs}$ is designed as
\begin{align}\label{eq33}
T_{xs}=&\frac{m}{\cos\alpha\cos\beta}\big(-k_{1v}\sqrt {\frac{L_v}{2}} \phi_{v1}+z_v\nonumber\\
&+\phi_{v3}(L_v,\phi_{v1})\big)\nonumber\\
\dot{z}_v=&-k_{2v}L_v\phi_{v2}
\end{align}
with $\phi_{v1}=\lceil {{S_V}} \rfloor^{1/2}+S_V$,
$\phi_{v2}=\phi_{v1}'\phi_{v1}=\frac{1}{2}\lceil {{S_V}} \rfloor^0+\frac{3}{2}\lceil {{S_V}} \rfloor^{1/2}+S_V$, and $\phi_{v3}(L_v,\phi_{v1})=-\frac{\dot{L}_v\phi_{v1}}{2L_v\phi'_{v1}}$. The derivative of $\phi_{v1} $ with respect to $S_V$ is $\phi '_{v1}=\frac{1}{2}\left|S_V\right|^{-\frac{1}{2}}+1$. To ease the notation, we denote $\phi_{v3}(L_v,\phi_{v1})=\phi_{v3}$. $k_{1v}$ and $k_{2v}$ are positive constants. $L_v$ is adaptive gain and adjusted via the following gain-adaptation law
\begin{equation}\label{eq34}
\left\{ {\begin{array}{*{20}{c}}
	{\begin{array}{*{20}{c}}
		{\begin{array}{*{20}{c}}
			{{L}_v = {L_{v0}} + \Delta {L}_v}\\
			{{\Delta {\dot L}_v} = - \lambda_v \left( t \right)\sign\left( {\bar{e}_{ v} } \right)}
			\end{array}}\\
		{\bar{e }_v = 0.5{L}_v - \frac{1}{l_v}\left| {{{\bar u}_{eqv}}} \right| - \varepsilon_v }
		\end{array}}\\
	{\lambda_v \left( t \right) = {\lambda _{v0}} + r_v\left( t \right)}
	\end{array}} \right.
\end{equation}
\begin{equation}\label{eq35}
\dot r_v\left( t \right)  = \left\{ {\begin{array}{*{20}{c}}
	{{\bar r_v \left| {\bar{e }_v} \right|\sign\left( {\left| {\bar{e }_v} \right| - e_b} \right),{\kern 1pt} {\kern 1pt} {\kern 1pt} r_v\left( t \right) > {r_{mv}}}}\\
	{{{r_{mv}},{\kern 1pt} {\kern 1pt} otherwise}}
	\end{array}} \right.,
\end{equation}
where ${L_{v0}}$, $\lambda_{v0} $, $\bar{r}_v$, $e _b$, and $r_{mv}$ are positive constants and $0 <  1/ l_v < 1$. $\varepsilon_v $ is small positive constant. $\bar{u}_{eqv}$ is the close approximation of equivalent control ${u}_{eqv}$, which is obtained by the following low-pass filter
\begin{equation}\label{eq36}
{\dot {\bar u}_{eqv}} = \frac{1}{\tau }\left[ \frac{k_{2v}L_v}{2}\lceil {{S_V}} \rfloor^0- {{\bar u}_{eqv}} \right],
\end{equation}
where time constant $\tau$ is set to be small enough so that $\bar{u}_{eqv}$ can be close to ${ u_{eqv}}$ arbitrarily \cite{Edwardsl2016Adaptive,Utkin2013Adaptive}.

Substituting control law \eqref{eq33} into \eqref{eq32} yields
\begin{align}\label{sudusystem}
\dot{S}_V=&-k_{1v}\sqrt {\frac{L_v}{2}} \phi_{v1}+\bar z_v+\phi_{v3}\nonumber\\
\dot{\bar z}_v=&-k_{2v}L_v\phi_{v2}+\dot{\Delta}_V
\end{align} 
with $\bar z_v=z_v+\Delta_V$.

\begin{remark}\label{rphi3}
	$\phi_{v3}$ is an adaptation term used to ensure the stability of system \eqref{sudusystem}. When $\phi_{v3}\ne 0$, $\phi_{v3}$ is an acceleration term, which improves the convergence speed. When $\dot{L}_v$ is equal to $0$ or $S_V$ converges to $0$ ($\phi_{v1}$ is also equal to $0$), $\phi_{v3}=0$ and the controller \eqref{eq33} reverts to the traditional generalized super-twisting structure.
\end{remark}

\begin{remark}\label{remark_bb}
	The gain-adaptation laws designed for attitude and airspeed subsystem are different (see \eqref{jeq24}-\eqref{eq25} and \eqref{eq34}-\eqref{eq35}). 
	For airspeed subsystem, because there only exists time-dependent disturbance $\Delta_V$, the two gains in \eqref{eq33} are both adjusted via the gain-adaptation law \eqref{eq34}-\eqref{eq35}. The parameter selection of this design
	is simple. Under the adjustment of \eqref{eq34}-\eqref{eq35}, the two gains of controller \eqref{eq33} are updated in real-time according to the change of disturbance, which can 
	attenuate the chattering efficiently.
\end{remark}

\begin{figure}[t]
	\centering
	\includegraphics[width=8cm]{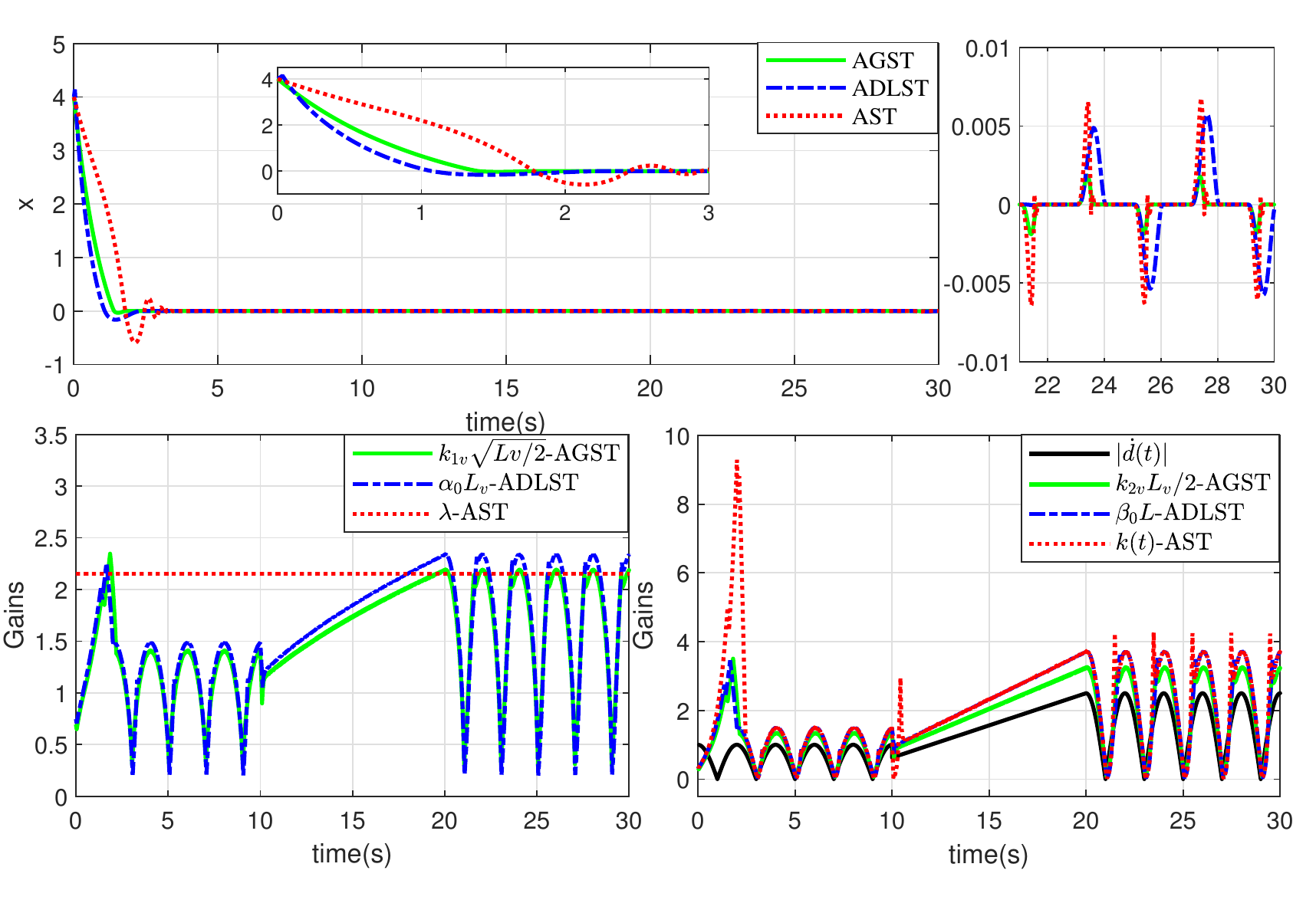}
	\caption{State $x$ and control gains.}
	\label{fig-chunsuanfa1}
\end{figure}

\begin{figure}[t]
	\centering
	\includegraphics[width=8cm]{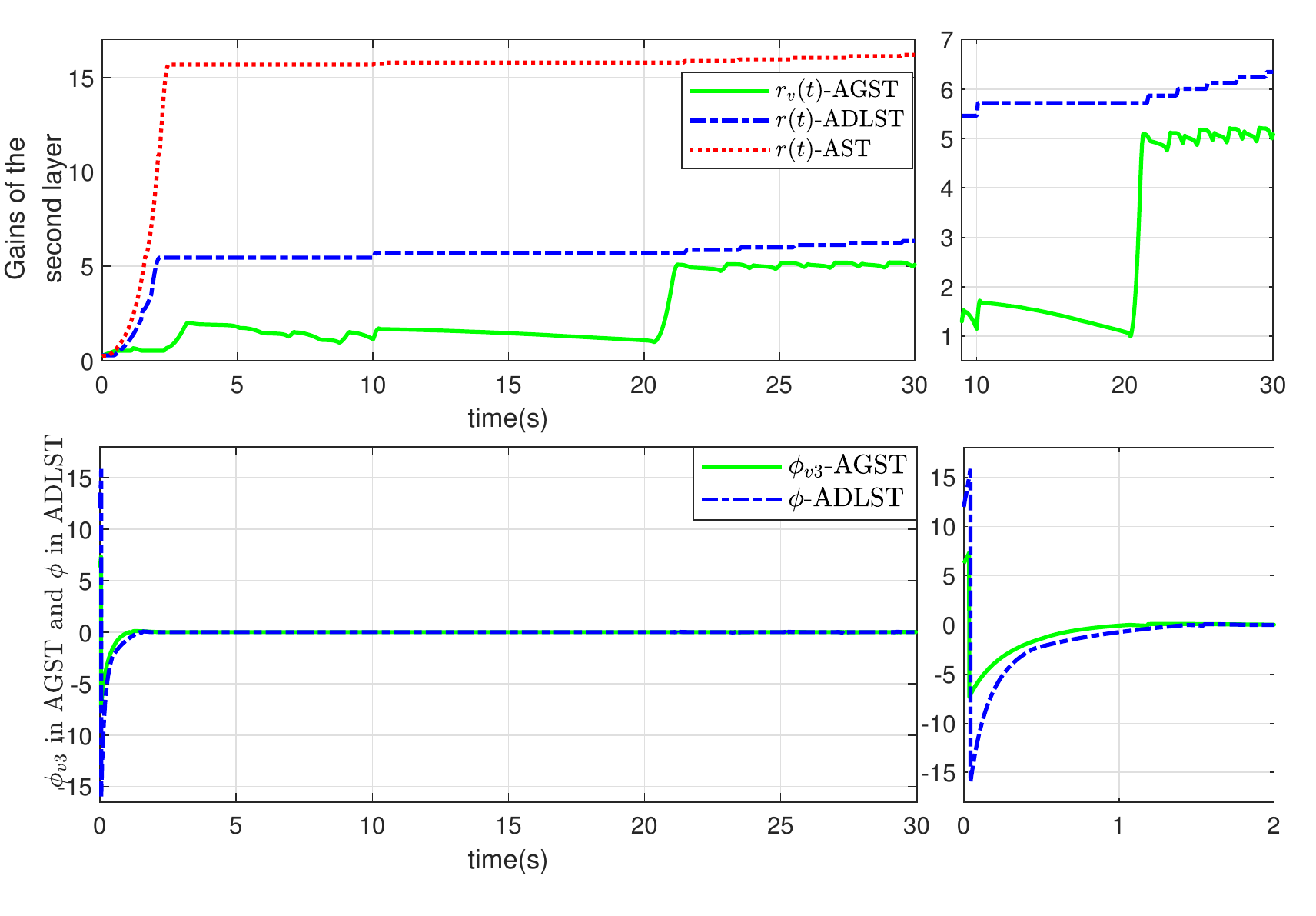}
	\caption{The second-layer gains in three control algorithms and $\phi_{v3}$, $\phi$ in AGST, ADLST.}
	\label{fig-chunsuanfa2}
\end{figure}

In comparison with the adaptive-gain super-twisting (AST) in \cite{Edwardsl2016Adaptive} and adaptive dual-layer super-twisting (ADLST) in \cite{Edwardsl2016lAdaptiveijc}, the proposed AGST has better control performance when facing unknown disturbance. The comparative results shown in Figs.~\ref{fig-chunsuanfa1} and \ref{fig-chunsuanfa2} illustrate this. For SISO system $\dot{x} = u+d(t)$, $x$ is the state and $u$ is the control input. The disturbance $d(t)$ is given as
$$d(t)=\left\{ {\begin{array}{*{20}{c}}
	{\frac{2\sin(0.5\pi t)}{\pi }, 0\le t<10}\\
	{\frac{3}{32}t^2-\frac{5}{4}t, 10\le t <20}\\
	{\frac{5\sin(0.5\pi t)}{\pi}, 20 \le t <30}
	\end{array}} \right..$$
The parameters setting of \eqref{eq34}-\eqref{eq35} is $k_{1v}=1.35$, $k_{2v}=1.26$, $L_{v0}=0.26$, $l_v=0.99$, $\varepsilon_v=0.05$, $\lambda_{v0}=0.38$, $\bar r_v=7$, $e_b=0.15$, and $r_{mv}=0.6$. The
parameters setting of ADLST in \cite{Edwardsl2016lAdaptiveijc} is $\alpha_0=1.35$, $\beta_0=1.26$, $a\beta_0=0.99$, $\epsilon=0.05$, $l_0=0.26$, $\gamma=7$, and $r_0=0.38$. The
parameters setting of gain-adaptation law of AST \cite{Edwardsl2016Adaptive} is same as that of ADLST and the fixed gain $\lambda$ is set as $2.2$. The specific definition of parameters in ADLST and AST can refer to \cite{Edwardsl2016lAdaptiveijc} and \cite{Edwardsl2016Adaptive}.

The results in Fig. \ref{fig-chunsuanfa1} show that the AGST has smaller overshoot and convergence time compared with other two algorithms. The convergence precision of AGST is also higher than the other two algorithms. 
$\frac{k_{2v}L_v}{2}$, $\beta_0L$, and $k(t)$ change with $|\dot d(t)|$ in real time, but the value of $\frac{k_{2v}L_v}{2}$ is smaller than $\beta_0L$ and $k(t)$ because the second-layer control gain $r_v(t)$ decreases after $10$s and fluctuates around $5$ after $20$s (see the top sub-graph of Fig. \ref{fig-chunsuanfa2}). The modified gain-adaptation law in \eqref{eq34} and \eqref{eq35} can avoid the overestimation of control gain $L_v$ and $r_v$ better, which will alleviate the chattering more effectively.
In addition, the results in Fig. \ref{fig-chunsuanfa2} indicate that the adaptation term $\phi_{v3}$ helps to achieve faster convergence speed in the initial phase, which confirm the descriptions in Remark \ref{rphi3}.

\begin{theorem}\label{theo2}
	For system \eqref{eq9} with Assumption \ref{assum3}, design the integral sliding manifold $S_V$ \eqref{tx} and sliding mode control law $T_{xs}$ \eqref{eq33} with gain-adaptation law \eqref{eq34}-\eqref{eq35}. Then, the system trajectory can reach the integral sliding manifold $S_V$ in finite time.	
\end{theorem}

\begin{pf}
The stability analysis of system \eqref{sudusystem}  
consists of two steps. 

\textit{Step 1:} Define $\bm X_v=[X_{v1},X_{v2}]^T=[\sqrt{\frac{L_v}{2}}\phi_{v1},\bar{ z}_v]^T$ and the derivative of $\bm X_v$ is
\begin{equation}\label{59}
\dot{\bm X}_v=\sqrt{L_v}\phi'_{v1}\big(\bm A_{v0}\bm X_v+\bm B_{v0}\tilde\Delta_V\big),
\end{equation} 
with $\tilde\Delta_V=\frac{\dot{\Delta}_V}{\sqrt{L_v}\phi'_{v1}}$, $\bm A_{v0}=\sqrt{2}\left[ {\begin{array}{*{20}{c}}
	{-\frac{k_{1v}}{2}}&{\frac{1}{2}}\\
	{-k_{2v}}&{0}
	\end{array}} \right]$, and $\bm B_{v0}=[0,1]^T$.

Similar to the proof process of Theorem \ref{theorem2}, suppose that there has been a gain-adaptation law to tune $L_v$ and ensure $L_v>\max\left\{ L_{v0}, 2\left|\dot{\Delta}_{V}\right| \right\}$. Further, from Assumption \ref{assum3}, we have $L_v>2\delta_{V0}$.
For system \eqref{59}, a Lyapunov function candidate is selected as
\begin{equation}\label{eqVV}
V_{v0}=\frac{1}{2}\bm X^T_v\bm P_v\bm X_v,
\end{equation}
where $\bm P_v=\left[ {\begin{array}{*{20}{c}}
	{\varpi_1}&{-1}\\
	{-1}&{\varpi_2}
	\end{array}} \right]$ is a positive definite matrix with $\varpi_1>0$ and $\varpi_1\varpi_2>1$. The goal of Step 1 is to obtain the stability condition of $k_{1v}$ and $k_{2v}$ such that the finite-time stability of system \eqref{sudusystem} can be obtained.

Taking the derivative of $V_{v0}$ with respect to time yields
\begin{align}\label{61}
\dot{V}_{v0}=&\frac{1}{2}\sqrt{L_v}\phi'_{v1}\big(\bm X^T_v\bm A^T_{v0}\bm P_v\bm X_v+\bm X^T_v\bm P_v\bm A_{v0}\bm X_v\nonumber\\
&+2\bm X^T_v\bm P_v\bm B_{v0}\tilde\Delta_{V}\big)\nonumber\\
\le&\frac{1}{2}\sqrt{L_v}\phi'_{v1}\big(\bm X^T_v\bm A^T_{v0}\bm P_v\bm X_v+\bm X^T_v\bm P_v\bm A_{v0}\bm X_v\nonumber\\
&+\bm X^T_v\bm P_v\bm B_{v0}\bm B^T_{v0}\bm P_v\bm X_v+\tilde\Delta^2_{V}\big).
\end{align}
From $\tilde\Delta_V=\frac{\dot{\Delta}_V}{\sqrt{L_v}(\frac{1}{2}\left|S_V\right|^{-\frac{1}{2}}+1)}$,
\begin{align}
\left|\tilde\Delta_V\right|
\le\frac{2\sqrt{2}\left|\dot{\Delta}_V\right|}{L_v}\left\|\bm X_{v}\right\|.
\end{align}
Because we have supposed that $L_v>\max\left\{ {{L_{v0}, 2\left|\dot\Delta_V\right|}} \right\}$,  $\left|\tilde\Delta_V\right|<\sqrt{2}\left\|\bm X_{v}\right\|$ and \eqref{61} is transformed into
\begin{align}\label{eqan}
\dot{V}_{v0}
\le&\frac{1}{2}\sqrt{L_v}\phi'_{v1}\bm X^T_v\big(\bm A^T_{v0}\bm P_v+\bm P_v\bm A_{v0}+\bm C^T_{v0}\bm C_{v0}\nonumber\\
&+\bm P_v\bm B_{v0}\bm B^T_{v0}\bm P_v\big)\bm X_v\nonumber\\
=&-\frac{1}{2}\sqrt{L_v}\phi'_{v1}\bm X^T_v\bm Q_{v0}\bm X_v,\nonumber
\end{align}
where $\bm C_{v0}=[\sqrt{2},0]^T$ and $\bm Q_{v0}=-\big(\bm A^T_{v0}\bm P_v+\bm P_v\bm A_{v0}+\bm C^T_{v0}\bm C_{v0}+\bm P_v\bm B_{v0}\bm B^T_{v0}\bm P_v\big)$. To ensure the negative definite of $\dot{V}_{v0}$, following condition has to be satisfied.
\begin{equation}\label{63}
\bm A^T_{v0}\bm P_v+\bm P_v\bm A_{v0}+\bm C^T_{v0}\bm C_{v0}+\bm P_v\bm B_{v0}\bm B^T_{v0}\bm P_v<0.
\end{equation}
Then, noting that $$\lambda_{\min}(\bm Q_{v0})\left\|\bm X_v\right\|^2\le\bm X^T_v\bm Q_{v0}\bm X_v\le\lambda_{\max}(\bm Q_{v0})\left\|\bm X_v\right\|^2,$$ $$\lambda_{\min}(\bm P_{v})\left\|\bm X_v\right\|^2\le\bm X^T_v\bm P_{v}\bm X_v\le\lambda_{\max}(\bm P_{v})\left\|\bm X_v\right\|^2,$$ 
we obtain
\begin{equation}\label{64}
\dot{V}_{v0}\le-\nu_{v1}{V}^{1/2}_{v0}-\nu_{v2}{V}_{v0}
\end{equation}
with $\nu_{v1}=\frac{L_{v0}\lambda_{\min}(\bm Q_{v0})\sqrt{\lambda_{\min}(\bm P_v)}}{4\lambda_{\max}(\bm P_v)}$ and $\nu_{v2}=\frac{\sqrt{L_{v0}}\lambda_{\min}(\bm Q_{v0})}{\lambda_{\max}(\bm P_v)}$. Based the Theorem 4.2 in \cite{Bhat2000Finite} and the assumption of $L_v$, it can be concluded that the system \eqref{59} is finite-time stable via selecting appropriate $k_{1v}$ and $k_{2v}$ to satisfy \eqref{63}.

\textit{Step 2:} In Step 2, the stability of adaptive gain $L_v$ is analyzed. Because the gain-adaptation law of $L_v$ is similar to that of $L$, the proving process of \eqref{eq34}-\eqref{eq35} is simply introduced in this step. 
On the reaching phase, $L_v$ is less than $2\delta_{V0}$ and $\bar e_v$ is negative. According to \eqref{eq34}-\eqref{eq35}, $L_v$ will increase until $L_v>2\delta_{V0}$. To examine the stability of $L_v$, choose the Lyapunov function as
\begin{equation}\label{L_vl}
V_{L_v}=\frac{1}{2}\bar e^2_v+\frac{1}{2\Gamma_v}(r_v-r^*_v)^2,
\end{equation}
where $r^*_v$ represents the upper bound of $r_v$ and $\Gamma_v>0$.
The derivative of $V_{L_v}$ is 
\begin{align}\label{dL_vl}
\dot{V}_{L_v}=&\bar e_v\dot{\bar e}_v+\frac{1}{\Gamma_v}(r_v-r^*_v)\dot{r}_v\nonumber\\
\le&-\frac{1}{2}\nu_{v_l}\left|\bar e_v\right|-\beta_{rv}\left|r_v-r^*_v\right|-\varsigma_{rv}\left|r_v-r^*_v\right|,
\end{align}
where $\nu_{v_l}=\lambda_{v0}+r_v-\frac{2\delta_{V1}}{l_v}$, $\beta_{rv}>0$, and $\varsigma_{rv}=-\beta_{rv}+\frac{1}{\Gamma_v}\bar r_v|\bar e_v|\sign(|\bar e_v|-e_b)$. To guarantee the positiveness of $\nu_{v_l}$, the condition  $\lambda_{v0}+r_v>\frac{2\delta_{V1}}{l_v}$ has to be satisfied. Based on the above condition, \eqref{dL_vl} is rewritten as $
\dot{V}_{L_v}\le -\bar \nu_{v}V^{1/2}_{Lv}-\varsigma_{rv}|r_v-r^*_v|
$
with $\bar \nu_{v}=\min\left\{ {\frac{1}{2}\nu_{v_l},\beta_{rv}} \right\}$. When $|\bar e_v|>e_b$, $r_v$ will increase and $\varsigma_{rv}$ is positive if
$$\Gamma_v<\frac{\bar r_ve_b}{\beta_{rv}}.$$
Then, inequality $\dot{V}_{L_v}\le -\bar \nu_{v}V^{1/2}_{Lv}$ holds, which illustrates that $\bar e_v$ can converge to the domain $\left|\bar e_v\right|<e_b$ in finite time \cite{Bhat2000Finite}. Similar to the analysis process of $e_\Delta$ in attitude subsystem, $\bar e_v$ could be sustained in a bigger domain, i.e., $\left|\bar e_v\right|<e_{b1}$ with $e_{b1}>e_{b}$, and $r_v$ has an upper bound $r^*_v$. 
Thereout, we can infer that the adaptive-gain $L_v$ is also bounded.

The proof of Theorem \ref{theo2} is completed.
\end{pf}

When the system trajectory reach and maintain on the sliding manifold $S_V$, with \eqref{eq32} and $\dot S_V=0$, the equivalent control $T_{xse}$ is deduced as
$$
T_{xse}=-\frac{m}{\cos\alpha\cos\beta}\Delta_V.
$$
Substitute $T_{xse}$ into \eqref{eq9} and then the sliding mode dynamic is written as
\begin{equation}\label{eq39}
\dot e_V=\frac{\cos\alpha \cos\beta T_{xa}-D}{m}-g_v-\dot V_d.
\end{equation}
 
From \eqref{eq39}, 
the effect of unknown disturbance $\Delta_V$ is completely compensated by the control law \eqref{eq33}. Next, the ADP approach can be employed to generate $T_{xa}$ such that the sliding mode dynamic \eqref{eq39} has a nearly optimal performance.
The design process of $T_{xa}$ is detailed in Subsection \ref{sec.c}

\subsection{Nearly optimal control design}\label{sec.c}

With the analysis in Subsection \ref{section3.2}, the tracking error systems \eqref{eq7} and \eqref{eq9} are converted to the equivalent sliding mode dynamics \eqref{peq27} and \eqref{eq39} when the system trajectories come to the ISMs.
By combining the results in Subsection \ref{section3.2}, the nearly optimal control laws $\bm M_a$ and $T_{xa}$ are designed via modified ADP approach in this subsection.

Define $\bm E_V=[\bm E^T, e_V]^T\in \mathbb{R}^{7}$, $\bm F_V(\bm{z_\Theta})=[\bm F^T(\bm{z_\Theta}),\\-\frac{D}{m}-g_v]^T\in \mathbb{R}^7$, $\bm X_d=[\bar{\bm \Theta}^T_d,\dot{V}_d]^T\in \mathbb{R}^7$, $\bm U_a=[\bm M_a^T, T_{xa}]^T\\ \in \mathbb{R}^4$, and 
$\bm G_V=[\bm{0}_{4\times3},\bm G_{V2}]^T\in \mathbb{R}^{7 \times 4}$ with 
$$\bm G_{V2}=\left[ {\begin{array}{*{20}{c}}
	{\bm{R_\Theta I}^{-1}}&{\bm 0_{3\times1}}\\
	{\bm 0_{1\times3}}&{\frac{\cos\alpha\cos\beta}{m}}
	\end{array}} \right]^T.$$
Then, \eqref{peq27} and \eqref{eq39} can be transformed into
\begin{equation}\label{eq40}
\dot{\bm E}_V=\bm F_V(\bm{z_\Theta})+\bm G_V\bm U_a-\bm X_d.
\end{equation}

\begin{assumption}\label{aG}
	From the engineering viewpoint, suppose that a positive constant $\delta_G$ can be found such that $\left\|\bm G_V\right\|\le\delta_G$.
\end{assumption}

\begin{remark}\label{remarkG}
	Through the expression of $\bm G_{V}$, we can observe that $\bm G_{V}$ is a matrix which related to the trigonometric function of $\bm \Theta$, $\alpha$, and $\beta$. Since $\theta$ cannot be equal to $ \pm \frac{\pi}{2}$ and then
	$\tan \theta$ is bounded,
	the assumption $\left\|\bm G_V\right\|\le\delta_G$ is reasonable.
\end{remark}

Based on \eqref{eq40}, the nearly optimal control law $\bm U_a$ is constructed. First, a infinite horizon performance index function $V_a(\bm E_V)$ is defined as
\begin{equation}\label{41}
V_a(\bm E_V)=\int_{t}^{\infty} {r_V(\bm E_V, \bm U_a)d\tau} 
\end{equation}
with $r_V(\bm E_V, \bm U_a)=\bar{\bm Q}_E(\bm {E}_V)+\bm U^T_a\bm R_u\bm U_a $
and $\bar{\bm Q}_E(\bm E_V)=\bm E^T_V{\bm Q}_E\bm E_V$. ${\bm Q}_E(\bm E_V)\in \mathbb{R}^{7\times7}$ is a positive-definite matrix.
$\bm R_u={diag}\left\{ {R_{u1},R_{u2},R_{u3},R_{u4}} \right\}\in \mathbb{R}^{4\times4}$ is a symmetric positive definite matrix. For $\bar{\bm Q}_E(\bm E_V)$, a small positive constant $b_E$ can be found such that $\bar{\bm Q}_E(\bm E_V)\ge b_E\left\|\bm E_V\right\|^2$.

Our aim is to obtain the optimal control law $\bm U_a$ to minimize the performance index function $V_a(\bm E_V)$ and stabilize the system \eqref{eq40}.

Associated with \eqref{eq40} and \eqref{41}, the Hamiltonian function is
\begin{align}\label{42}
H(\bm E_V, \bm U_a, \nabla V_a)=&\nabla V_a\big(\bm F_V(\bm{z_\Theta})+\bm G_V\bm U_a-\bm X_d\big)\nonumber\\
&+\bar{\bm Q}_E(\bm E_V)+\bm U^T_a\bm R_u\bm U_a
\end{align}
with $\nabla V_a=\frac{\partial V_a}{\partial \bm E_V}$. The optimal control law is $\bm U^*_a$ and the optimal performance index is 
\begin{equation}\label{vv*}
V^*_a(\bm E_V)=\int_{t}^{\infty} {r_V(\bm E_V, \bm U^*_a)d\tau}.
\end{equation}

According to \eqref{vv*}, the Hamilton-Jacobi-Bellman (HJB) equation is
\begin{align}\label{eq44}
H(\bm E_V, \bm U^*_a, \nabla V^*_a)=&\nabla V^{*T}_a\big(\bm F_V(\bm{z_\Theta})+\bm G_V\bm U^*_a-\bm X_d\big)\nonumber\\
&+\bar{\bm Q}_E(\bm E_V)+\bm U^{*T}_a\bm R_u\bm U^*_a=0
\end{align}
with $\nabla V^*_a=\frac{\partial V^*_a}{\partial \bm E_V}$.
Then, $\bm U^*_a$ can be yielded as
\begin{equation}\label{eq45}
\bm U^*_a=-\frac{1}{2}\bm R^{-1}_u\bm G^{T}_V\nabla V^*_a=[\bm M^{*T}_a, T^*_{xa}]^T,
\end{equation}
and \eqref{eq44} is transformed into
\begin{align}\label{eq46}
H(\bm E_V, \bm U^*_a, \nabla V^*_a)=&\nabla V^{*T}_a\big(\bm F_V(\bm{z_\Theta})-\bm X_d\big)+\bar{\bm Q}_E(\bm E_V)\nonumber\\
&-\frac{1}{4}\nabla V^{*T}_a\bm G_V\bm R^{-1}_u\bm G^{T}_V\nabla V^{*}_a=0.
\end{align}
Through \eqref{eq45} and \eqref{eq46}, the gradient $\nabla V^{*}_a$ is achieved. Next, the optimal control law is obtained through $\nabla V^{*}_a$. 
Nevertheless, due to the strong nonlinearity, it is difficult or even impossible to solve the nonlinear HJB equation via analytical methods. Hence, a modified ADP approach based on the actor-critic (AC) structure is proposed to overcome the above difficulty.

\paragraph{Design of Critic NN and Actor NN:}
In the AC structure, the optimal control law $\bm U_a^*$ and the optimal performance index $V_a^*(\bm E_V)$ are approximated by actor neural network (ANN) and critic neural network (CNN), respectively.

In our work, $V^{*}_a(\bm E_V)$ is divided into two parts, which is shown as
\begin{align}\label{eq47}
V^{*}_a(\bm E_V)=&\beta_w\left\|\bm E_V\right\|^2-\beta_w\left\|\bm E_V\right\|^2+V^{*}_a(\bm E_V)\nonumber\\
=&\beta_w\left\|\bm E_V\right\|^2+V^{*}_{a2}(\bm E_V),
\end{align}
where $\beta_w>0$ and $V^{*}_{a2}(\bm E_V)=-\beta_w\left\|\bm E_V\right\|^2+V^{*}_a(\bm E_V)$. According to the Weierstrass high-order approximation theorem \cite{Fan2016Adaptive}, there exist a neural network (NN) such that $V^{*}_{a2}(\bm E_V)$ is approximated as
\begin{equation}\label{jVa2}
V^{*}_{a2}(\bm E_V)=\bm W^{*T}_c\bm \sigma_{w}(\bm E_V)+\delta_w,
\end{equation}
where $\bm W^*_c=[W^*_{c1},...,W^*_{cN}]^T$ is the optimal weight vector, $N$ represents the number of neurons. ${\bm \sigma _w}\left( {\bm{E }_V} \right)=[\sigma_{w1},\sigma_{w2},...,\\\sigma_{wN}]^T$ is the suitable activation function vector and each elements in ${\bm \sigma _w}\left( {\bm{E }_V} \right)$ is selected to be linearly independent. ${\delta _w}$ is the approximation error. Substituting \eqref{jVa2} into \eqref{eq47} yields
\begin{equation}\label{eq48}
V^{*}_a(\bm E_V)=\beta_w\left\|\bm E_V\right\|^2+\bm W^{*T}_c\bm \sigma_{w}(\bm E_V)+\delta_w
\end{equation}
and $\nabla V^{*}_a$ is
\begin{equation}\label{eq49}
\nabla V^{*}_a=2\beta_w\bm E_V+\nabla \bm \sigma^T_{w}\bm W^*_c+\nabla \delta_w
\end{equation}
where $\nabla \bm \sigma_w=\frac{\partial \bm \sigma_{w}(\bm E_V)}{\partial \bm E_V}$ and $\nabla \delta_w=\frac{\partial\delta_w}{\partial \bm E_V}$.

\begin{assumption}\label{youjie}
	According to the universal approximation principle, the approximation error $\delta_w$ and its gradient are bounded, i.e., $\left\|\delta_w\right\|\le\bar\delta_w$ and $\left\|\nabla\delta_w\right\|\le\bar\delta_{w0}$ with $\bar\delta_w>0$, $\bar\delta_{w0}>0$. Besides, activation function vector $\bm \sigma_w (\bm E)$ and its gradient are also bounded, i.e., $\left\|\bm \sigma_w(\bm E_V)\right\|\le\bar\delta_\sigma$ and $\left\|\nabla\bm \sigma_w\right\|\le\bar\delta_{\sigma0}$ with $\bar\delta_\sigma>0$, $\bar\delta_{\sigma0}>0$ \cite{Zhang2018Optimal,Guo2018Reinforcement}. 
\end{assumption}

Substituting \eqref{eq49} into \eqref{eq45}, rewrite the optimal control law $\bm U^*_a$ as
\begin{equation}\label{eq50}
\bm U^*_a=-\frac{1}{2}\bm R^{-1}_u\bm G^{T}_V(2\beta_w\bm E_V+\nabla \bm \sigma^T_{w}\bm W^*_c)+\delta_{wu}
\end{equation}
where $\delta_{wu}$ is the approximation error of the optimal control law. $\delta_{wu}$ is relevant to $\delta_\omega$ and satisfies $\left\|\delta_{wu}\right\|\le \bar \delta_{wu}, \bar \delta_{wu}>0$.

According to \eqref{eq49} and \eqref{eq50}, HJB equation is derived as
\begin{align}\label{eq51}
&H(\bm E_V, \bm U^*_a, \nabla V^*_a)=(2\beta_w\bm E_V+\nabla \bm \sigma^T_{w}\bm W^*_c)\big(\bm F_V(\bm{z_\Theta})\nonumber\\
&-\bm X_d\big)+\bar{\bm Q}_E(\bm E_V)-\frac{1}{4}\big(2\beta_w\bm E_V+\nabla \bm \sigma^T_{w}\bm W^*_c\big)^T \bm G_V
\nonumber\\
&\cdot\bm R^{-1}_u\bm G^{T}_V\big(2\beta_w\bm E_V+\nabla \bm \sigma^T_{w}\bm W^*_c\big)-\delta_{HJB}=0
\end{align}
where $\delta_{HJB}$ is the residual error and related to the approximation error $\delta_w$ and $\delta_{wu}$. A positive constant can be found such that $\left\|\delta_{HJB}\right\|\le\bar\delta_{h}$.

Sine the ideal weight vector is unknown, the CNN and ANN are constructed as follows.
\begin{empheq}{align}
\hat V_a({\bm E}_V)&=\beta_w\left\|{\bm E}_V\right\|^2+\hat{\bm W}^{T}_c\bm \sigma_{w}({\bm E}_V), \label{eq52}\\
\hat{\bm U}_a=&-\frac{1}{2}\bm R^{-1}_u\bm G^{T}_V(2\beta_w{\bm E}_V+\nabla {\bm \sigma}^T_{w}\hat{\bm W}_a)\nonumber\\
=&\ [\bm M^T_a, T_{xa}]^T, \label{eq53}
\end{empheq}
and $\nabla \hat{V}_a $ is
\begin{equation}\label{eq54}
\nabla \hat{V}_a=2\beta_w{\bm E}_V+\nabla{\bm \sigma}^T_{w}\hat{\bm W}^{T}_c,
\end{equation}
where $\nabla{\bm \sigma}^T_{w}=\frac{\partial \bm \sigma({\bm E}_V)}{\partial {\bm E}_V}$. $\hat{\bm W}_c$ and $\hat{\bm W}_a$ are the weight vectors of CNN and ANN, respectively. It is obvious that $\hat{\bm W}_c$ and $\hat{\bm W}_a$ are used to approximate the same ideal $\bm W^{*}_c$. 

From \eqref{eq53} and \eqref{eq54}, the HJB equation is given by
\begin{align}\label{eq55}
&\hat H({\bm E}_V, \hat{\bm U}_a, \nabla \hat{V}_a)=(2\beta_w{\bm E}_V+\nabla {\bm \sigma}^T_{w}\hat{\bm W}_c)\big(\bm F_V({\bm z}_\Theta)\nonumber\\
&+\bm G_V \hat{\bm U}_a-\bm X_d\big)+\bar{\bm Q}_E({\bm E}_V)+\hat{\bm U}^T_a\bm R_u\hat{\bm U}_a,
\end{align}
and the Bellman residual error $\Delta_B$ is
\begin{align}\label{eq56}
\Delta_B=&\hat H({\bm E}_V, \hat{\bm U}_a, \nabla \hat{V}_a)-H(\bm E_V, \bm U^*_a, \nabla V^*_a)\nonumber\\
=&\hat H({\bm E}_V, \hat{\bm U}_a, \nabla \hat{V}_a).
\end{align}
To minimize the Bellman residual error $\Delta_B$, 
define a objective function as ${E_c}=\frac{1}{2}\Delta _B ^T{\Delta _B }.$
$\hat{\bm W_c}$ and $\hat{\bm W_a}$ are updated to minimize $E_c$. Based on the gradient descent method, the updating law of $\hat{\bm W_c}$ is obtained as
\begin{align}\label{eq57}
\dot{\hat {\bm W}}_c=&-c_0\frac{m_w}{(1+m^T_wm_w)^2}\big[(2\beta_w{\bm E}_V+\nabla {\bm \sigma}^T_{w}\hat{\bm W}_c)\big(\bm F_V({\bm z}_\Theta)\nonumber\\
&+\bm G_V \hat{\bm U}_a-\bm X_d\big)+\bar{\bm Q}_E({\bm E}_V)+\hat{\bm U}^T_a\bm R_u\hat{\bm U}_a\big].	
\end{align}
According to the stability analysis, the adaptive weight tuning law for ${\hat {\bm W}_{a}}$ is derived as 
\begin{align}\label{eq58}
\dot{\hat {\bm W}}_a=&-a_0\big[(\bm \Gamma_a\hat{\bm W}_a-\bm \Gamma_bm_{1w}\hat{\bm W}_c)-\frac{1}{4}\hat{\bm W}_a\nabla{\bm\sigma}_w\bm A\nabla{\bm\sigma}^T_w\bar{m}_w\nonumber\\
&\cdot \hat{\bm W}_c\big]+\frac{a_0}{2}\Pi(\hat{\bm U}_a)\nabla{\bm\sigma}_w\bm A\nabla\Psi,
\end{align}
where $m_w=\nabla{\bm \sigma}^T_w\big(\bm F_V({\bm z}_\Theta)+\bm G_V \hat{\bm U}_a-\bm X_d\big)$, $\bar m_w=\frac{m_w}{(1+m^T_wm_w)^2}$, $m_{1w}=\frac{m_w}{(1+m^T_wm_w)}$, and $\bm A=\bm G_V\bm R^{-1}_u\bm G^T_V$. $c_0>0$ and $a_0>0$ are the learning rates of CNN and ANN, respectively. $\bm \Gamma_a$ and $\bm \Gamma_b$ are tuning matrix and vector to stabilize the system. The expression of $\Pi(\hat{\bm U}_a)$ is shown as
\begin{equation}\label{eq59}
\Pi(\hat{\bm U}_a)  = \left\{ {\begin{array}{*{20}{c}}
	{0, if \ \nabla\Psi^T(\bm F_V({\bm z}_{\bm \Theta})+\bm G_V \hat{\bm U}_a-\bm X_d)}<0\\
	{1, otherwise}
	\end{array}} \right..
\end{equation}
The term $\frac{a_0}{2}\Pi(\hat{\bm U}_a)\nabla\bm\sigma_w\bm A\nabla\Psi$ is an adaptation term, which ensures the stability of system \eqref{eq40} during the training process and relaxes the requirement of initial stabilizing control. The definition and properties of function $\Psi$ are given in Assumption \ref{psi}. 
	
\begin{figure}[t]
	\centering
	\includegraphics[width=8cm]{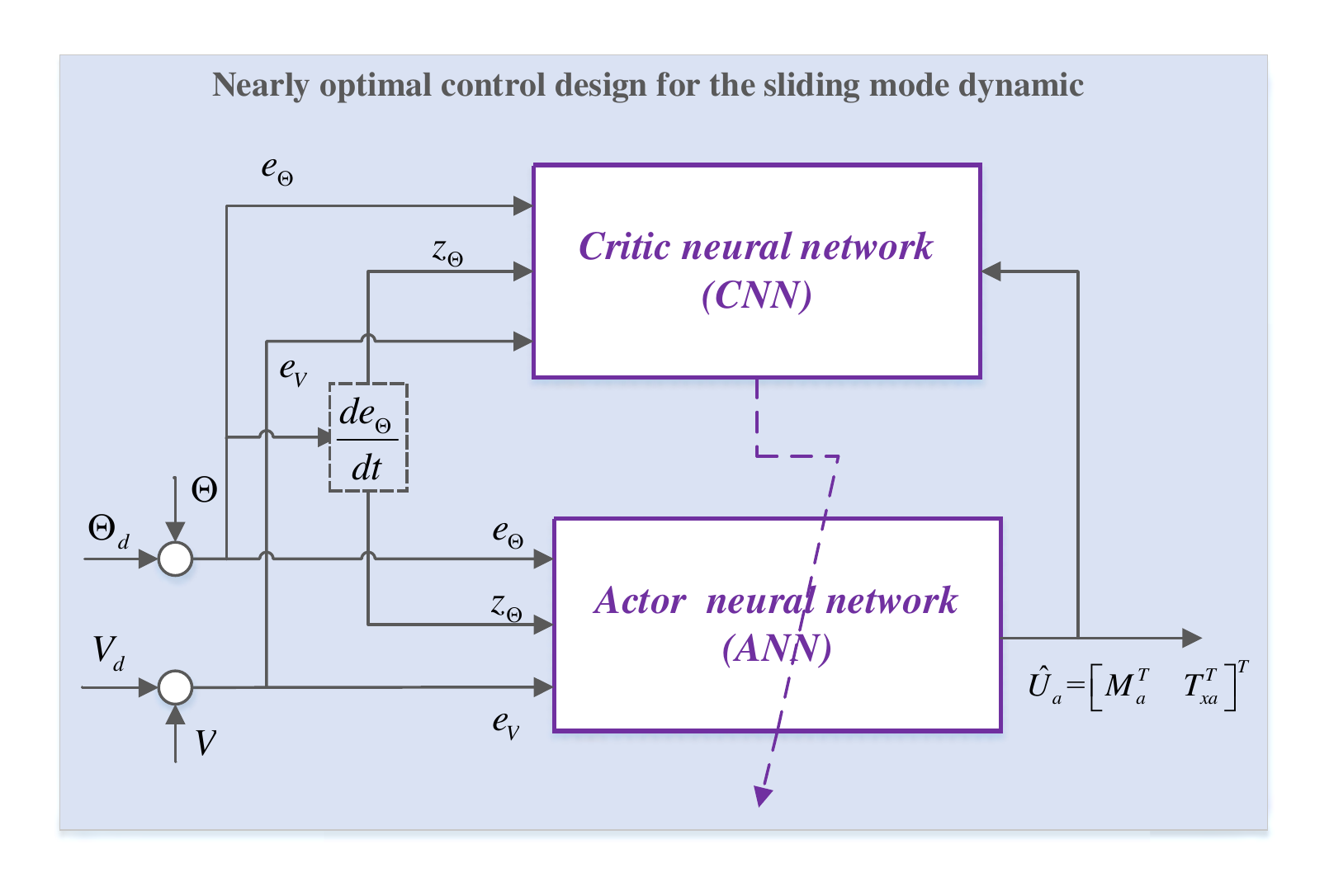}
	\caption{Structure of the nearly optimal control scheme.}
	\label{fig-adp}
\end{figure}

The structure of the nearly optimal control scheme, which is based on the modified ADP approach, is given
in Fig.~\ref{fig-adp}.

\begin{assumption}\label{psi}\cite{Guo2018Reinforcement,Liu2017Adaptivedynamic}
	Suppose that $\Psi$ is a continuously differentiable radially unbounded Lyapunov function and its gradient along system \eqref{eq40} is $\nabla \Psi$. There exists a condition that $\dot\Psi=\nabla \Psi(\bm F_V(\bm{z_\Theta})+\bm G_V\bm U^*_a-\bm X_d)<0$ holds. Moreover, we can find a positive-definite matrix $\bm Q_{\Psi}\in \mathbb{R}^{7\times7}$ such that $\dot\Psi=-\nabla \Psi\bm Q_{\Psi}\nabla \Psi<0$.	
\end{assumption}

\begin{theorem}\label{theorem3} 
	Consider system \eqref{eq40} with Assumptions \ref{aG}-\ref{psi}. Design the CNN and ANN as \eqref{eq52} and \eqref{eq53}. Then, let the weight tuning laws of two NNs be described by \eqref{eq57} and \eqref{eq58}, respectively. Select relative parameters appropriately, then the tracking error vector $\bm{E}_V=[\bm e_\Theta,\bm e_V]^T$, weight estimations errors ${\tilde {\bm W}_{ c}}=\bm{W_c}^* - {\hat {\bm W}_{c}}$ and ${\tilde {\bm W}_{ a}}=\bm{W_c}^* - {\hat {\bm W}_{ a}}$ are UUB.
\end{theorem}

\begin{pf}

Design the Lyapunov function candidate as follows.
\begin{align}\label{eq83}
V_A=
\Psi+V^*_a+L_c+L_a,
\end{align}	
where $L_c=\frac{1}{2}{\tilde {\bm W}^T_{ c}}c^{-1}_0{\tilde {\bm W}_{ c}}$ and $L_a=\frac{1}{2}{\tilde {\bm W}^T_{ a}}a^{-1}_0{\tilde {\bm W}_{ a}}$. $\Psi$ is the continuous Lyapunov function given in Assumption \ref{psi}.  Taking derivative of $V_A$, we have
\begin{align}\label{eq84}
\dot{V}_A=&\dot\Psi+\dot V^*_a+\dot{L}_c+\dot{L}_a.
\end{align}

First, $\dot V^*_a$ is analyzed. Based on the designed $\hat{\bm U}_a$, the derivative of $V^*_a$ is
\begin{align}\label{Vad}
\dot V^*_a=&\nabla V^*_a(\bm F_V(\bm{z_\Theta})+\bm G_V\hat{\bm U}_a-\bm X_d)\nonumber\\
=&\delta_{VA}+\bm A^*\bm F_V(\bm{z_\Theta})-\bm A^*\bm X_d\nonumber\\
&-\frac{1}{2}\bm A^*\bm G_V\bm R^{-1}_u\bm G^{T}_V\bm A_a,
\end{align}
where $\bm A^*=2\beta_w\bm E_V+\nabla \bm \sigma^T_{w}\bm W^*_c$ and $\bm A_a=2\beta_w{\bm E}_V+\nabla \bm {\sigma}^T_{w}\hat{\bm W}_a$. $\delta_{VA}=\nabla \delta_w\big(\bm F_V(\bm{z_\Theta})
-\frac{1}{2}\bm G_V\bm R^{-1}_u\bm G^{T}_V(2\beta_w{\bm E}_V+\nabla {\bm \sigma}^T_{w}\hat{\bm W}_a)-\bm X_d\big)$, which satisfies $\left\|\delta_{VA}\right\|\le \delta_{VA1}\left\|{\bm E}_V\right\|+\delta_{VA2}$.
Noting the fact that 
\begin{align}
\bm A^*(\bm F_V(\bm{z_\Theta})-\bm X_d)=&-\bar{\bm {Q}}_E(\bm E_V)+\frac{1}{4}\bm A^*\bm G_V\bm R^{-1}_u\bm G^T_V\bm A^*\nonumber\\
&+\delta_{HJB}\nonumber
\end{align}
and $\bar{\bm Q}_E(\bm E_V)\ge b_E\left\|\bm E_V\right\|^2$. \eqref{Vad} can be transformed into
\begin{align}\label{eq86}
\dot V^*_a\le&\delta_{VA}+\delta_{HJB}-b_E\left\|\bm E_V\right\|^2+\frac{1}{2}\bm A^*\bm G_V\bm R^{-1}\bm G^T_V \tilde{\bm A}_a,
\end{align}
where 
$\tilde{\bm A}_a=\bm A^*-\bm A_a=\nabla{\bm \sigma}^T_w\tilde{\bm W}_a.$  

Considering $\dot{\tilde{\bm W}}_c=-\dot{\hat{\bm W}}_c$, the derivative of $L_c$ is
\begin{align}
\dot{L}_c=&\tilde{\bm W}_c\frac{m_w}{(1+m^T_wm_w)^2}\big[\bm A_c\big(\bm F_V({\bm z}_\Theta)+\bm G_V \hat{\bm U}_a-\bm X_d\big)\nonumber\\
&+\bar{\bm Q}_E({\bm E}_V)+\hat{\bm U}^T_a\bm R_u\hat{\bm U}_a\big],\nonumber
\end{align}
where $\bm A_c=2\beta_w{\bm E}_V+\nabla{\bm \sigma}^T_w\hat{\bm W}_c$. From \eqref{eq51}, there exists
\begin{align}
\bar{\bm Q}_E({\bm E}_V)=&\delta_{HJB}-\bm A^*(\bm F_V(\bm {z_\Theta})-\bm X_d)\nonumber\\
&+\frac{1}{4}\bm A^{*T}\bm{G}_V\bm R^{-1}_u\bm{G}^T_V\bm A^*.
\end{align} 
$\dot{L}_c$ is rewritten as
\begin{align}\label{88}
\dot{L}_c
\le&\tilde{\bm W}^T_c\frac{m_w}{(1+m^T_wm_w)^2}\big[-{m^T_w}\tilde{\bm W}_c+\delta_{HJB}\nonumber\\
&+\frac{1}{4}\tilde{\bm A}_a\bm G_V\bm R^{-1}_u\bm{G}^T_V\tilde{\bm A}_a\big],
\end{align}
where $\tilde{\bm A}_c=\bm A^*-\bm A_c=\nabla{\bm \sigma}^T_w\tilde{\bm W}_c$.

Next, considering $\dot{\tilde{\bm W}}_a=-\dot{\hat{\bm W}}_a$, the derivative of $L_a$ is
\begin{align}\label{89}
\dot L_a
=&-\tilde{\bm W}_a\bm \Gamma_a\tilde{\bm W}_a+\tilde{\bm W}_a\bm \Gamma_a\bm W^*_c-\tilde{\bm W}_a\bm \Gamma_bm_{1w}{\bm W}^*_c\nonumber\\
&+\tilde{\bm W}_a\bm \Gamma_bm_{1w}\tilde{\bm W}_c-\frac{1}{4}\tilde{\bm W}_a(\bm W^*_c-\tilde{\bm W}_a)\nabla{\bm\sigma}_w\bm A\nabla{\bm\sigma}^T_w\nonumber\\
&\cdot\bar{m}_w (\bm W^*_c-\tilde{\bm W}_c)
-\frac{\tilde{\bm W}_a}{2}\Pi(\hat{\bm U}_a)\nabla{\bm\sigma}_w\bm A\nabla\Psi.
\end{align}

According to the results in \eqref{eq86}, \eqref{88}, and \eqref{89}, it is easy to show that
\begin{align}\label{90}
\dot V_A\le\bar \Psi-\bm X^T_a\bm\Xi_1\bm X_a+\left\|\bm\Xi_2\right\|\left\|\bm X_a\right\|+\bm\Xi_3,
\end{align}
where $\bm X_a=[\bm E^T_V,m_{1w}\tilde{\bm W}^T_c,\tilde{\bm W}^T_a]^T$, $\bar \Psi=\nabla \Psi^T(\bm F_V({\bm z}_{\bm \Theta})+\bm G_V\hat{\bm U}_a-\bm X_d) -\frac{\tilde{\bm W}_a}{2}\Pi(\hat{\bm U}_a)\nabla{\bm\sigma}_w\bm A\nabla\Psi$. $\bm \Xi_1$, $\bm \Xi_2$, and $\bm \Xi_3$ are shown as follows.
$$\bm \Xi_1=\left[ {\begin{array}{*{20}{c}}
	{\chi_1}&{\frac{\chi_5}{2}}&{\frac{\chi_4}{2}}\\
	{\frac{\chi_5}{2}}&{\chi_2}&{\frac{\chi_6}{2}}\\
	{\frac{\chi_4}{2}}&{\frac{\chi_6}{2}}&{\chi_3}
	\end{array}} \right],
\bm\Xi_2=[\eta_1, \eta_2, \eta_3]^T,$$
$$\bm\Xi_3=\delta_{VA2}+\delta_{HJB}.$$
$$\chi_1=b_E,\quad \chi_2=\bm I,\quad \chi_3=\bm \Gamma_a-\frac{1}{4}\nabla\bm \sigma^T_w\bm A\nabla{\bm \sigma}_w\bar{m}_\omega\bm W^*_c$$
$$\chi_4=-\beta_w\bm A\nabla{\bm \sigma}^T_w, \quad\chi_5=0$$
$$\chi_6=-\bm\Gamma_b-\frac{\nabla{\bm \sigma}^T_w\bm A\nabla{\bm \sigma}_w\bm W^{*}_c\bar {m}_\omega}{4}$$
$$\eta_1=\delta_{VA1},\quad \eta_2=\frac{\delta_{HJB}}{1+m^T_wm_w},$$
\begin{align}
\eta_3=&\bm \Gamma_a\bm W^*_c-\bm \Gamma_bm_{1w}\bm W^*_c-\frac{\bm W^{*T}_c\nabla{\bm \sigma}_w\bm A\nabla{\bm \sigma}^T_w\bm W^{*}_c\bar m_w}{4}\nonumber\\
&+\frac{\nabla{\bm \sigma}_w\bm A\nabla{\bm \sigma}^T_w\bm W^{*}_c}{2}.\nonumber
\end{align}

For matrix $\bm \Xi_1$, $\bm \Gamma_a$ and $\bm \Gamma_b$ are selected appropriately to ensure the positive definite of $\bm \Xi_1$. Besides, there exist two positive constants $\delta_{\Xi_2}$ and $\delta_{\Xi_3}$, such that $\left\|\bm \Xi_2\right\|\le\delta_{\Xi_2}$ and $\left\|\bm \Xi_3\right\|\le\delta_{\Xi_3}$.

On basis of the result in \eqref{89}, the following classified discussion is carried out.

\textit{Case 1 $(\Pi(\hat{\bm U}_a)=0)$:} In this case, we have $$\nabla\Psi^T(\bm F_V({\bm z}_{\bm \Theta})+\bm G_V \hat{\bm U}_a-\bm X_d)<0.$$ 
According to \cite{Liu2017Adaptivedynamic}, there exists a positive constant $\eta_\Psi$ satisfying $\nabla\Psi^T(\bm F_V({\bm z}_{\bm \Theta})+\bm G_V \hat{\bm U}_a-\bm X_d)<-\eta_\Psi\left\|\nabla\Psi\right\|<0$. \eqref{89} is rewritten as
\begin{align}\label{eq90}
\dot V_A
\le&-\eta_\Psi\left\|\nabla\Psi\right\|-\lambda_{\min}(\bm\Xi_1)\big(\left\|\bm X_a\right\|-\frac{\delta_{\Xi_2}}{2\lambda_{\min}(\bm\Xi_1)}\big)^2\nonumber\\
&+\frac{\delta^2_{\Xi_2}}{4\lambda_{\min}(\bm\Xi_1)}+\delta_{\Xi_3}.
\end{align}
By \eqref{eq90}, $\dot V_A<0$ when
\begin{align}
\left\|\nabla\Psi\right\|>\frac{\delta^2_{\Xi_2}}{4\eta_\Psi\lambda^2_{\min}(\bm\Xi_1)}+\frac{\delta_{\Xi_3}}{\eta_\Psi}
\end{align} 
or
\begin{align}
\left\|\bm X_a\right\|>\frac{\delta_{\Xi_2}+\sqrt {\delta^2_{\Xi_2}+4\lambda_{\min}(\bm\Xi_1)\delta_{\Xi_3}} }{2\lambda_{\min}(\bm\Xi_1)}.
\end{align}

\textit{Case 2 ($(\Pi(\hat{\bm U}_a)=1)$):} In this case, $\bar \Psi$ is transformed into
\begin{align}
\bar \Psi
=&\nabla \Psi^T(\bm F_V({\bm z}_{\bm \Theta})+\bm G_V{\bm U}^*_a-\bm X_d)
-\nabla \Psi^T \bm G_V\delta_{\omega u}\nonumber\\
\le&-\eta_x\lambda_{\min}(\bm Q_{\Psi})\left\|\nabla\Psi\right\|^2+\delta_{y1}\left\|\nabla\Psi\right\|\nonumber
\end{align}
with $\eta_x\in(0,1)$ and $\delta_{y1}=\left\|\bm G_V\right\|\bar{\delta}_{\omega u}$.
Then, \eqref{90} is rewritten as
\begin{align}\label{93}
\dot{V}_A\le&-\eta_x\lambda_{\min}(\bm Q_{\Psi})\left\|\nabla\Psi\right\|^2+\delta_{y1}\left\|\nabla\Psi\right\|\nonumber\\
&-\delta_{y0}\left\|\bm X_a\right\|^2+\delta_{\Xi_2}\left\|\bm X_a\right\|+\delta_{\Xi_3}\nonumber\\
=&-\eta_x\lambda_{\min}(\bm Q_{\Psi})\big(\left\|\nabla\Psi\right\|-\frac{\delta_{y1}}{2\eta_x\lambda_{\min}(\bm Q_{\Psi})}\big)^2\nonumber\\
&-\delta_{y0}(\left\|\bm X_a\right\|-\frac{\delta_{\Xi_2}}{2\delta_{y0}})^2+\bm \Xi_4,
\end{align}
where $\delta_{y0}=\lambda_{\min}(\bm\Xi_1)$ and $\bm \Xi_4=\frac{\delta^2_{\Xi_2}}{4\delta_{y0}}+\delta_{\Xi_3}+\frac{\delta^2_{y1}}{4\eta_x\lambda_{\min}(\bm Q_{\Psi})}$. $\dot{V}_A$ is negative when
\begin{equation}\label{94}
\left\|\nabla\Psi\right\|>\sqrt{\frac{\bm \Xi_4}{\eta_x\lambda_{\min}(\bm Q_{\Psi})}}+\frac{\delta_{y1}}{2\eta_x\lambda_{\min}(\bm Q_{\Psi})}
\end{equation}
or
\begin{equation}\label{95}
\left\|\bm X_a\right\|>\sqrt{\frac{\bm \Xi_4}{\delta_{y0}}}+\frac{\delta_{\Xi_2}}{2\delta_{y0}}.
\end{equation}

Through the above analysis, $\nabla\Psi$, $\bm E_V$, $\tilde{\bm W_c}$, and $\tilde{\bm W_a}$ are UUB. Thus, the tracking error $\bm e_{\bm \Theta}$ and $e_V$ in attitude and airspeed subsystem are UUB.

The proof of Theorem \ref{theorem3} is completed.
	
\end{pf}

\subsection{Design procedure of the ADP-ASMC scheme}\label{sec.d}
For system \eqref{eq5} and \eqref{eq8} satisfying Assumption \ref{assum2}-\ref{youjie}, the design procedures of the proposed ADP-ASMC scheme is given as follows.

\textit{Step 1:} Utilize the AC structure-based ADP approach to approximately solve the HJB equation  \eqref{eq46} of sliding mode dynamics \eqref{eq40}. Through \eqref{eq57} and \eqref{eq58}, the weight estimation $\hat{\bm W}_c$ and $\hat{\bm W}_a$ can be obtained. Then, the nearly optimal control laws are given as
$$\left[ {\begin{array}{*{20}{c}}
	{\bm M_a}\\
	{T_{xa}}
	\end{array}} \right]=\hat{\bm U}_a=-\frac{1}{2}\bm R^{-1}_u\bm G^{T}_V(2\beta_w{\bm E}_V+\nabla {\bm \sigma}^T_{w}\hat{\bm W}_a).$$

\textit{Step 2:} On the basis of the nearly optimal control laws in Step 1, ISMs in two subsystems are formulated as
$$\bm S =\bm {z_\Theta} -\int_{0}^{t}{ \bm {R_\Theta I}^{-1}\bm M_a- {{{\bm{\ddot \Theta }}}_{\bm{d}}}}d\tau_t, $$
and
$$S_V=e_V(t)-\int_{0}^{t}{-g_v-\dot {V}_d+\frac{{\cos \alpha \cos \beta{T_{xa}}-D }}{m}}d\tau_t.$$
For the above ISMs, the sliding mode control laws are designed as
$$\left\{ {\begin{array}{*{20}{c}}
	{\bm M_s =\bm I \bm {R_\Theta}^{-1}\big(-k_{1} \bm \Phi_1 ({\bm S})+\bm z_1-\bm{G({\bm z}_\Theta)}\big)}\\
	{\dot{\bm z}_1=-k_{20}L\bm \Phi_2 ({\bm S})}
	\end{array}} \right.$$
with gain-adaptation law \eqref{jeq24}-\eqref{eq25}, and
$$\left\{ {\begin{array}{*{20}{c}}
	{T_{xs}=\frac{m}{\cos\alpha\cos\beta}\left(-k_{1v}\sqrt {\frac{L_v}{2}} \phi_{v1}+z_v+\phi_{v3}(L_v,\phi_{v1})\right)}\\
	{\dot{z}_v=-k_{2v}L_v\phi_{v2}}
	\end{array}} \right. $$
with gain-adaptation law \eqref{eq34}-\eqref{eq35}, respectively.

\textit{Step 3:} Following Step 1 and Step 2, set the control laws for fixed-wing UAV as 
$$\left[ {\begin{array}{*{20}{c}}
	{\bm M}\\
	{T_x}
	\end{array}} \right]=\left[ {\begin{array}{*{20}{c}}
	{\bm M_s}\\
	{T_{xs}}
	\end{array}} \right]+\left[ {\begin{array}{*{20}{c}}
	{\bm M_a}\\
	{T_{xa}}
	\end{array}} \right].$$ 

\begin{figure}[t]
	\centering
	\includegraphics[width=8.5 cm]{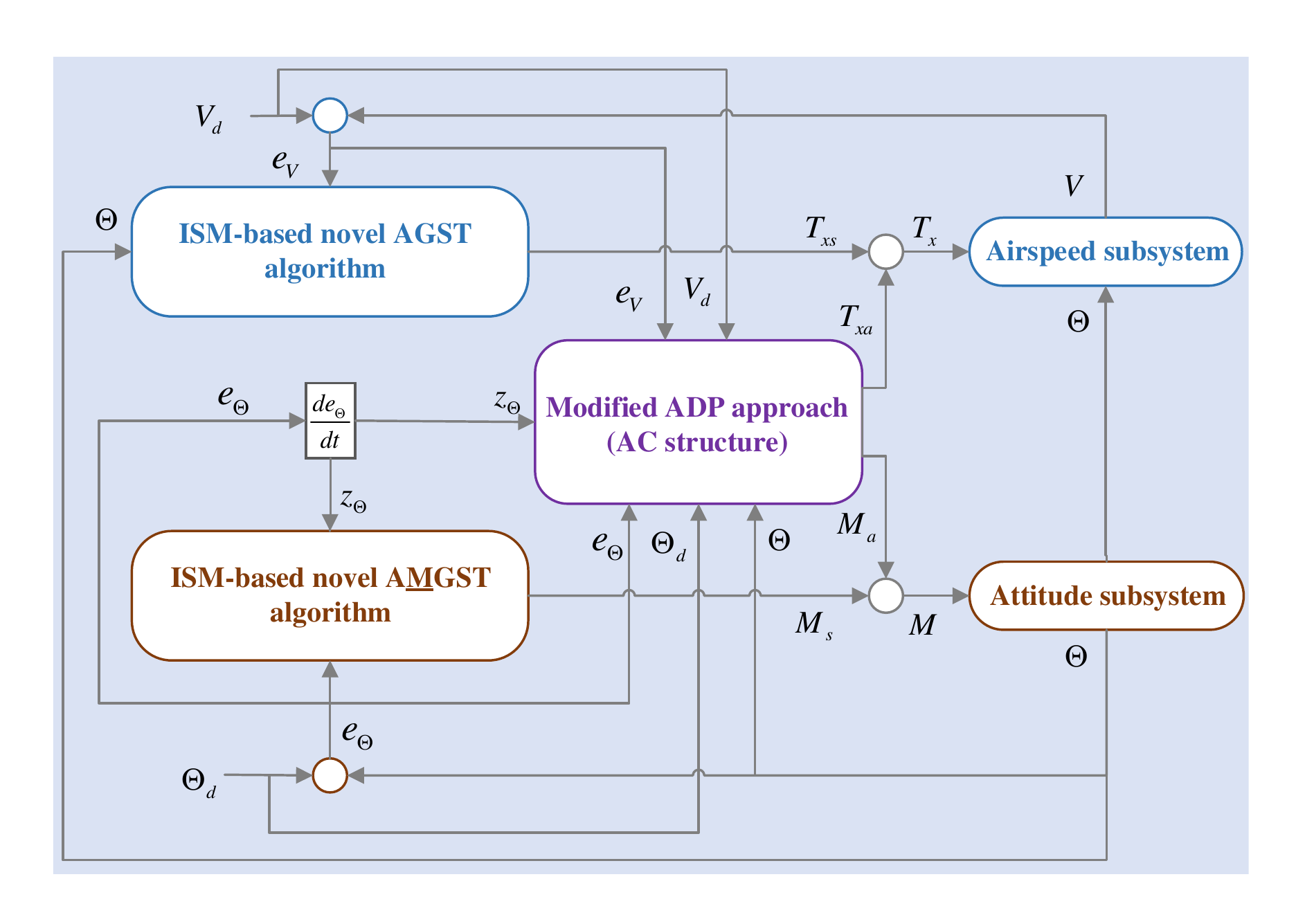}
	\caption{Schematic diagram of ADP-ASMC scheme.}
	\label{fig-juti}
\end{figure}

To illustrate the overall control architecture clearly in our work,
the schematic diagram of the proposed ADP-ASMC scheme is shown in Fig. \ref{fig-juti}

\section{NUMERICAL SIMULATIONS}\label{section5}

In this section, numerical simulations are conducted for demonstrating the efficiency of the proposed ADP-ASMC scheme. The simulation scenarios, results, and analysis are all given.


The parameters of fixed-wing UAV used in our work refer to \cite{Castaneda2017Extended} and \cite{Stevens2015Aircraft}. The initial conditions of fixed-wing UAV are: $\bm \Theta_0 =[5.8, -11.5, 11.5]^T $(deg), $\bm \omega_0 = [0.58, 1.15,\\ 1.72]^T$(deg/s), and $V_0=
0.4$(m/s). The nominal inertia matrix $\bm I$ is:
$I_{xx} = 0.5528$, $I_{yy} = 0.6335$, $I_{zz} = 1.0783$, and $I_{xz} = 0.0015$. The matched disturbance
$\bm \Delta \bm d_m$, unmatched disturbance $\bm \Delta \bm d_u$,  and external disturbance $\Delta_V (t)$ are given as follows.
$$\bm \Delta \bm d_m=\left\{ {\begin{array}{*{20}{c}}
	{\left[0;0;0\right]^T, 0<t<5}\\
	{\left[1.5\sin(\pi t/17);0.8\sin(\pi t/15);1.1\sin(\pi t/16)\right]^T}
	\end{array}} \right.,$$
$$\bm \Delta \bm d_u=[2.1\sin(\pi t/19);2.1\sin(\pi t/19);2.1\sin(\pi t/19)],$$
$$\Delta_V=\left\{ {\begin{array}{*{20}{c}}
	{0, 0<t<6}\\
	{5\sin(0.2t), t\ge 6}
	\end{array}} \right..$$
The numerical simulations are implemented on MATLAB/$\\$SIMULINK environment and the integration step is specified as 1ms.

In this part, the proposed attitude control scheme is compared with two control methods, i,e., the linear sliding surface-based adaptive second order sliding mode (LSS-ASOSM) in \cite{Castaneda2017Extended} and adaptive continuous twisting algorithm (ACTA) in \cite{Moreno2016Adaptive}. The LSS-ASOSM is designed on the basis of linear sliding surface and adaptive second order sliding mode. In addition, the designed airspeed control scheme is contrasted with ASOSM in \cite{Castaneda2017Extended} and fast terminal sliding mode surfac-based generalized super-twisting (FTSM-GST) method. It should be noted that the FTSM-GST method is derived from \cite{Dong2017Integrated}. 
The FTSM surface is given as
\begin{equation}
S_{vf}=e_V+k_s\int_{0}^{t}{\lceil {{e_V}} \rfloor^{\gamma_{f1}}+\lceil {{e_V}} \rfloor^{\gamma_{f2}}}d\tau_t\nonumber
\end{equation}
with $k_s>0$, $\gamma_{f1}\ge1$, and $\gamma_{f2}\in (0,1)$. The thrust based on FTSM-GST method is given as
\begin{align}
T_x=&\frac{m}{\cos\alpha\cos\beta}\big[\frac{D}{m}+g_v+\dot V_d-ks(\lceil {{e_V}} \rfloor^{\gamma_{f1}}+\lceil {{e_V}} \rfloor^{\gamma_{f2}})\nonumber\\
&-k_{1f}\phi_{f1}+z_f\big]\nonumber\\
\dot {z}_f=&-k_{2f}\phi_{f2},
\end{align}
where $\phi_{f1}=\lceil {{S_{vf}}} \rfloor^{\frac{1}{2}}+S_{vf}$ and $\phi_{f2}=\frac{1}{2}\lceil {{S_{vf}}} \rfloor^0+\frac{3}{2}\lceil {{S_{vf}}} \rfloor^{\frac{1}{2}}+S_{vf}$.

The parameter setting of the ISM-based AMGST controller (for attitude subsystem) and ISM-based AGST controller (for airspeed subsystem) are given in Table~\ref{table1}.
\begin{table} 
	\centering
	\caption{The parameters of ISM-based AMGST and ISM-based AGST.} 
	\begin{tabular}{cc}
		\toprule
		Controller & Parameters   \\
		\midrule
		ISM-based AMGST    & \tabincell{c}{$\kappa_1=8,\ \kappa_0=0.2,$ \\ $L_0=0.3,\ al=0.99 ,$ \\ $\varepsilon=0.01,\ \lambda_0=0.01,\ \bar r=10,$ \\ $ \bar e=0.1,\ r_m=0.6.$}  \\
		\hline 
		ISM-based AGST    & \tabincell{c}{ $k_{1v}=5, k_{2v}=3, L_{v0}=0.55,$\\$ l_v=0.99, \lambda_{v0}=0.01,\varepsilon_v =0.05$\\$e_b=0.3,\ \bar {r}_v=5,\ r_{mv}=0.5.$ }  \\
		\bottomrule
	\end{tabular}
	\label{table1}
\end{table}

\begin{table} 
	\centering
	\caption{The parameters of LSS-ASOSM, ACTA, ASOSM and FTSM-GST.} 
	\begin{tabular}{cc}
		\toprule
		Controller & Parameters   \\
		\midrule
		LSS-ASOSM    & \tabincell{c}{$\lambda_l=1,\ k=15,$ \\ $\mu=0.005,\ K_{\min}=0.8,$ \\ $\epsilon=1.35.$}  \\
		\hline 
		ACTA    & \tabincell{c}{ $l=5, k_{1t} = 1.1; k_{2t} = 1.1,$ \\ $k_{3t}=1.2, k_{4t}=1.2.$ }  \\
		\hline 
		ASOSM    &\tabincell{c}{ ${k_V} = 12, {\mu} = 0.01,$\\ ${K_{V\min}} = 0.8, {\epsilon_V} = 1.$ } \\
		\hline 
		FTSM-GST    &\tabincell{c}{ ${\gamma_{f1}} = 1.2, {\gamma_{f2}} = 0.88, k_s=1.5,$ \\${k_{1f}} = 4, {k_{2f}} = 1.5.$ } \\
		\bottomrule
	\end{tabular}
	\label{table2}
\end{table}

The control parameters of nearly optimal controller are provided as follows. The matrices $\bm Q_E$ and $\bm R_u$ are set as: $\bm Q_E=1.5\times\bm I_7$ and $\bm R_u=diag\left\{ {1.2,1.23,1,2.2} \right\}$. The initial weight values in CNN and ANN are selected in the range $(0,2]$ randomly. The number of neurons in CNN and ANN is $N =35$ and the activation function vector is selected as
\begin{align}
\bm \sigma_w(\hat{\bm E}_V)=\big[&e^2_{\Theta 1},e_{\Theta 1}e_{\Theta 2},e^2_{\Theta 2},e_{\Theta 1}e_{\Theta 3},e^2_{\Theta 3},e_{\Theta 2}e_{\Theta 3},z^2_{\Theta 1},\nonumber\\
&z_{\Theta 1}z_{\Theta 2},z^2_{\Theta 2},z_{\Theta 1}z_{\Theta 3}, z^2_{\Theta 3}, z_{\Theta 2}z_{\Theta 3},e^3_{\Theta 1}z_{\Theta 1},\nonumber\\
&e^3_{\Theta 2} z_{\Theta 2},e^3_{\Theta 3}z_{\Theta 3},e_{\Theta 1} z_{\Theta 1}z_{\Theta 2},e_{\Theta 2} z_{\Theta 2}z_{\Theta 3},\nonumber\\
&e_{\Theta 3}z_{\Theta 3}z_{\Theta 1},e_{\Theta 1}z_{\Theta 2},e_{\Theta 1}z_{\Theta 3},e_{\Theta 2}z_{\Theta 1},e_{\Theta 2}z_{\Theta 3},\nonumber\\
&e_{\Theta 3}z_{\Theta 1},e_{\Theta 3} z_{\Theta 2},z^3_{\Theta 1}e_{\Theta 3}e_{\Theta 2},z^3_{\Theta 2}e_{\Theta 1}e_{\Theta 3},\nonumber\\
&z^3_{\Theta 3}e_{\Theta 1}e_{\Theta 2},e_{\Theta 1}z^3_{\Theta 1},e_{\Theta 2}z^3_{\Theta 2},e_{\Theta 3}z^3_{\Theta 3}, e^2_V, \nonumber\\
&e_Ve_{\Theta 1},e_Ve_{\Theta 2},e_Ve_{\Theta 1}^3,e_Ve_{\Theta 2}^3\big]^T.\nonumber
\end{align}

In Table~\ref{table2}, the control parameters of the ACTA, ASOSM, LSS-ASOSM, and FTSM-GST are specified . The specific definition of parameters in four comparative control methods can refer to \cite{Castaneda2017Extended}, \cite{Moreno2016Adaptive}, and \cite{Dong2017Integrated}.

\begin{figure}[t]
	\centering
	\includegraphics[width=7.8cm]{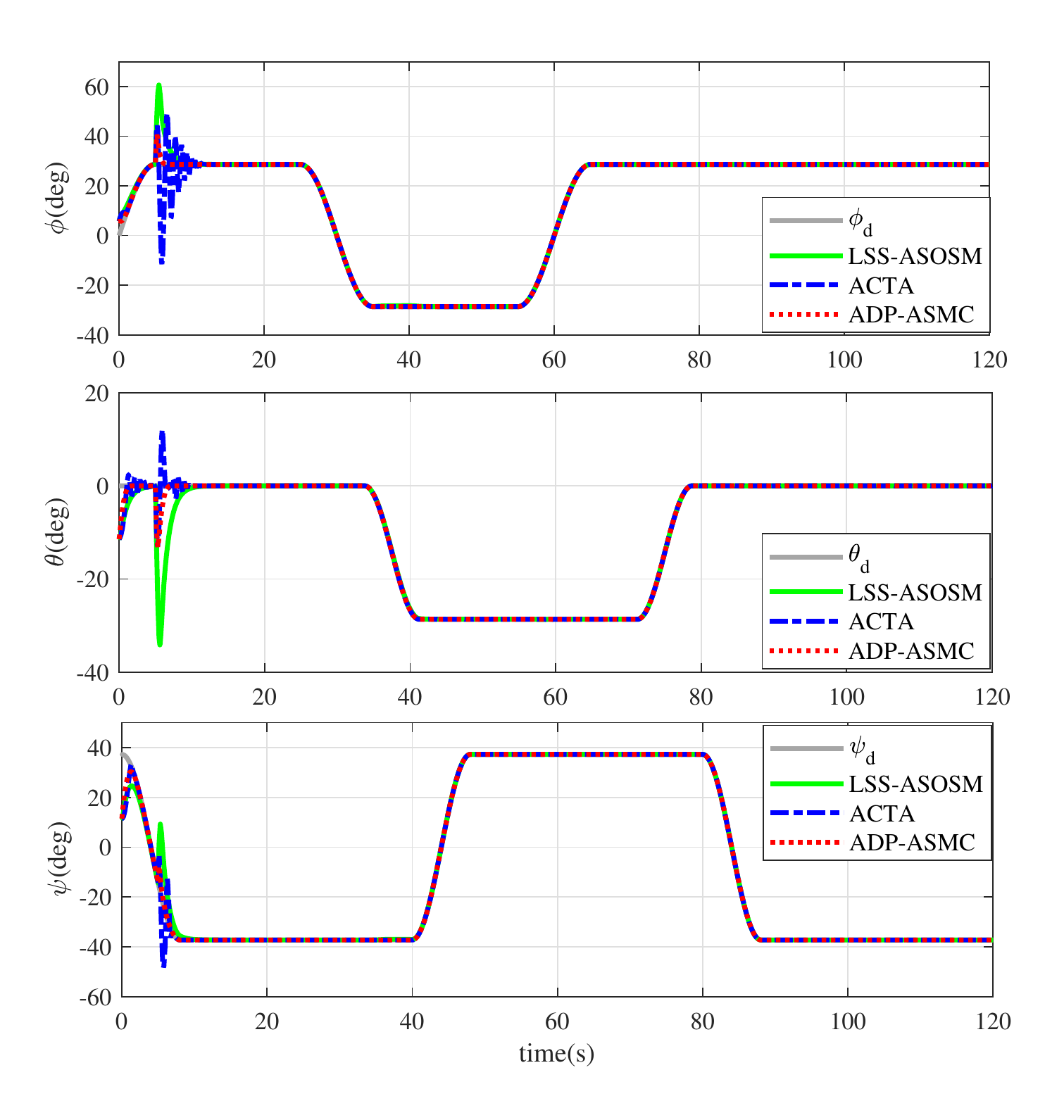}
	\caption{Tracking performance of attitude subsystem under LSS-ASOSM, ACTA, and ADP-ASMC in 0-120s.}
	\label{zitaigenzong}
\end{figure}

\begin{figure}[!htb]
	\centering
	\includegraphics[width=7.8cm]{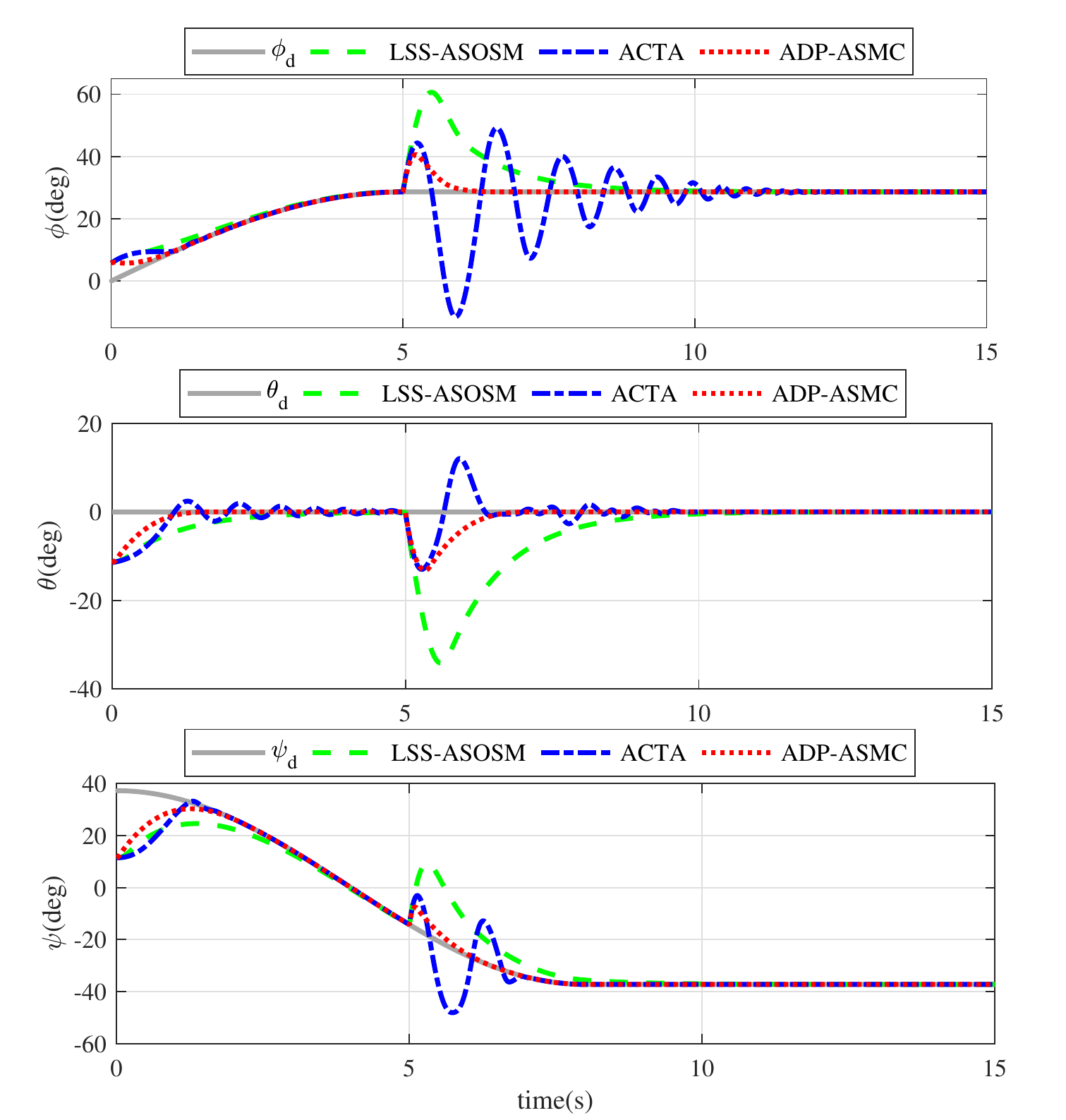}
	\caption{Tracking performance of attitude subsystem under LSS-ASOSM, ACTA, and ADP-ASMC in 0-15s.}
	\label{zitaigenzong2}
\end{figure} 

The tracking performance under ACTA, LSS-ASOSM, and ADP-ASMC are shown in Figs.~\ref{zitaigenzong} and \ref{zitaigenzong2}. The attitude angles $\phi$, $\theta$, and $\psi$ can track the reference command under the three control methods. From the local figures in Fig. \ref{zitaigenzong2}, the ADP-ASMC performs better than the other two control methods when the unmatched disturbance $\bm{\Delta d}_u$ is imposed at $t = 5s$.

\begin{figure}[t]
	\centering
	\includegraphics[width=7.8cm]{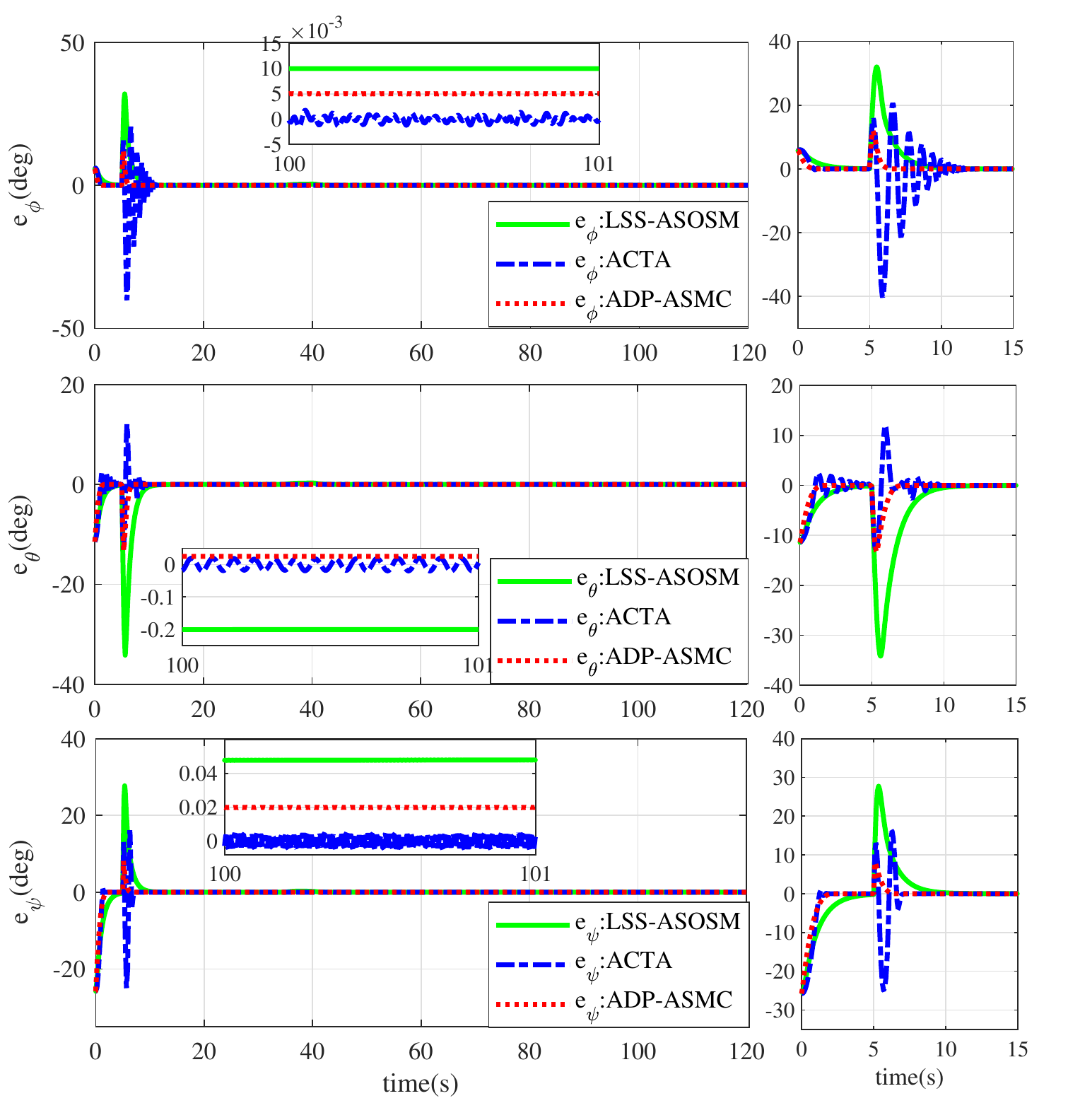}
	\caption{Tracking errors of $\phi$, $\theta$, $\psi$ under LSS-ASOSM, ACTA, and ADP-ASMC.}
	\label{zitaiwucha}
\end{figure}

\begin{figure}[!htb]
	\centering
	\includegraphics[width=7.8cm]{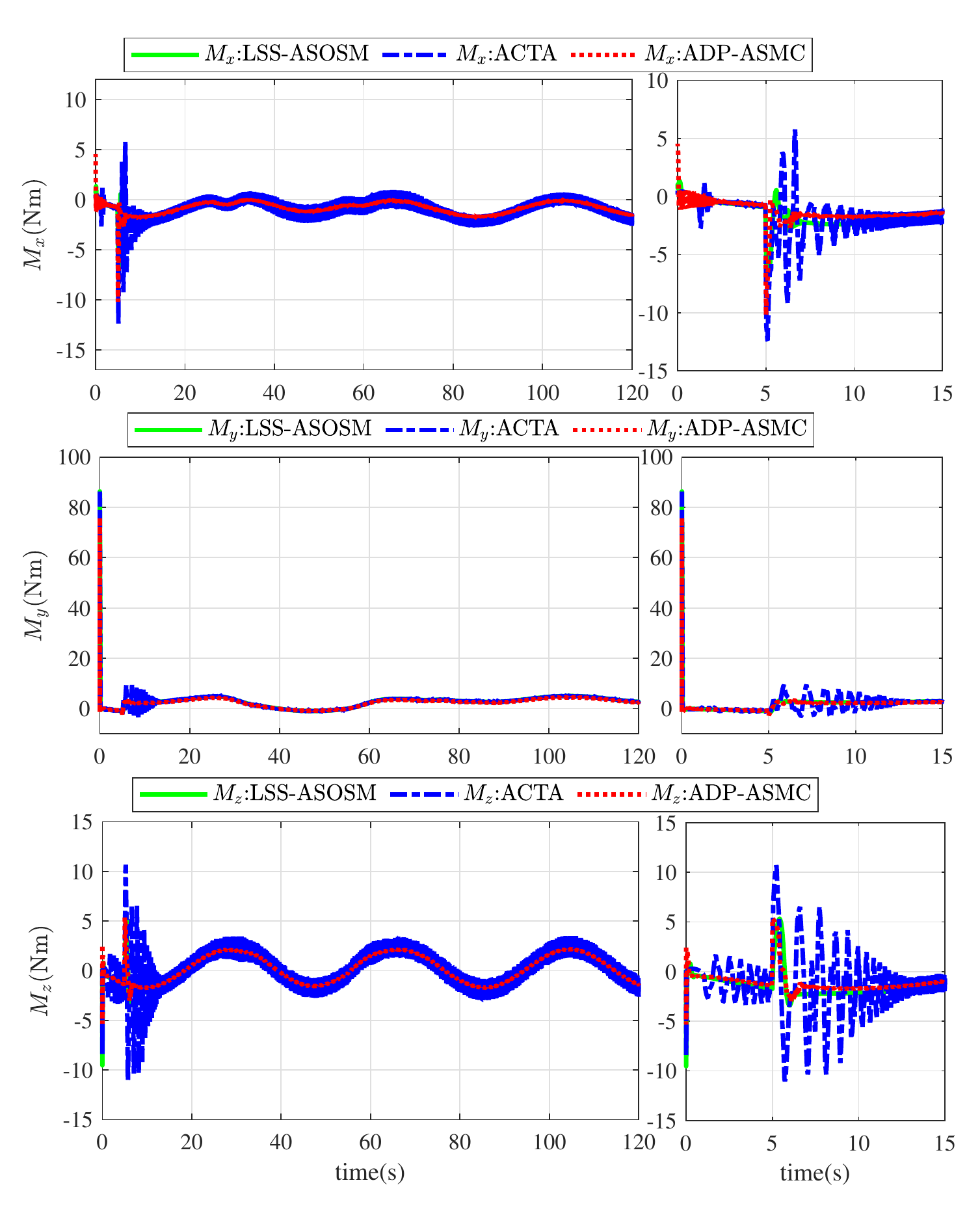}
	\caption{Control moments $M_x$, $M_y$, $M_z$ of LSS-ASOSM, ACTA, and ADP-ASMC.}
	\label{liju}
\end{figure}

\begin{figure}[!htb]
	\centering
	\includegraphics[width=7.8cm]{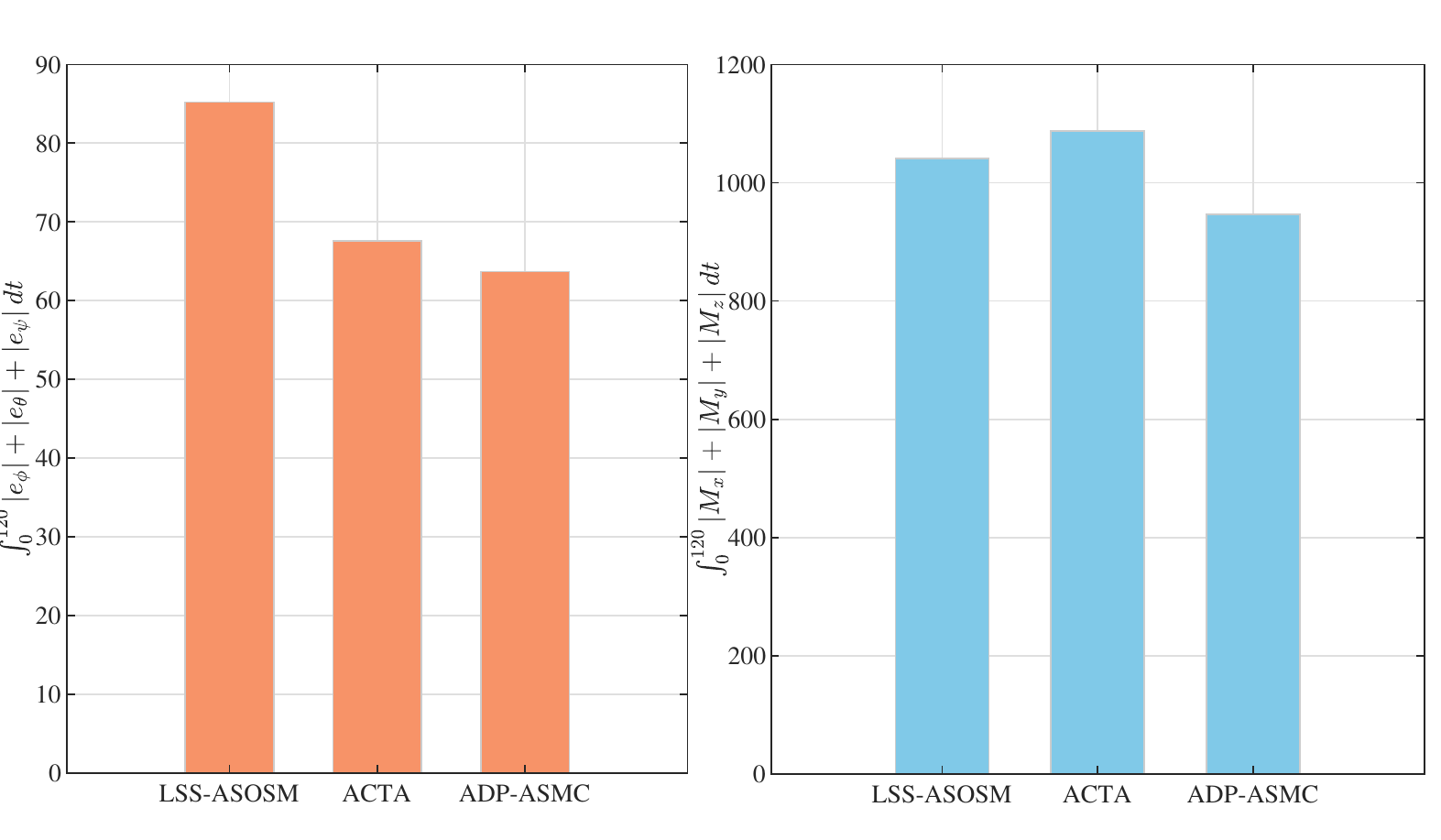}
	\caption{Performance index of three control schemes.}
	\label{performance}
\end{figure}

In Fig.~\ref{zitaiwucha}, tracking errors $e_\phi$, $e_\theta$, $e_\psi$ are presented. Fig.~\ref{liju} shows the control moments $M_x$, $M_y$, $M_z$. From Fig. \ref{zitaiwucha}, we can see that the tracking errors can be stabilized through three control methods. 
From the simulation results in Fig. \ref{liju}, the chattering amplitude in ADP-ASMC are much smaller than that of the other two control methods, which is beneficial for the actuators. The reason lies in two aspects: i). The modified gain-adaptation laws \eqref{jeq24}-\eqref{eq25} help avoid the overestimation of control gains. ii). The modified ADP approach can generate nearly optimal control moments $\bm M_a$, which optimizes the energy consumption to some extent. 
The performance index comparisons among three control methods presented in Fig.~\ref{performance} has also illustrated above point. The integral absolute error (IAE) $\int_{0}^{120}(|e_\phi|+|e_\theta|+|e_\psi|)dt$ and integral absolute control moment (IACM) $\int_{0}^{120}(|M_x|+|M_y|+|M_z|)dt$ are used as performance indexes. It is obvious that the proposed ADP-ASMC outperforms the existing LSS-ASOSM and ACTA. ADP-ASMC achieves the satisfactory tracking performance and the least energy consumption.

\begin{figure}[!htb]
	\centering
	\includegraphics[width=7.8cm]{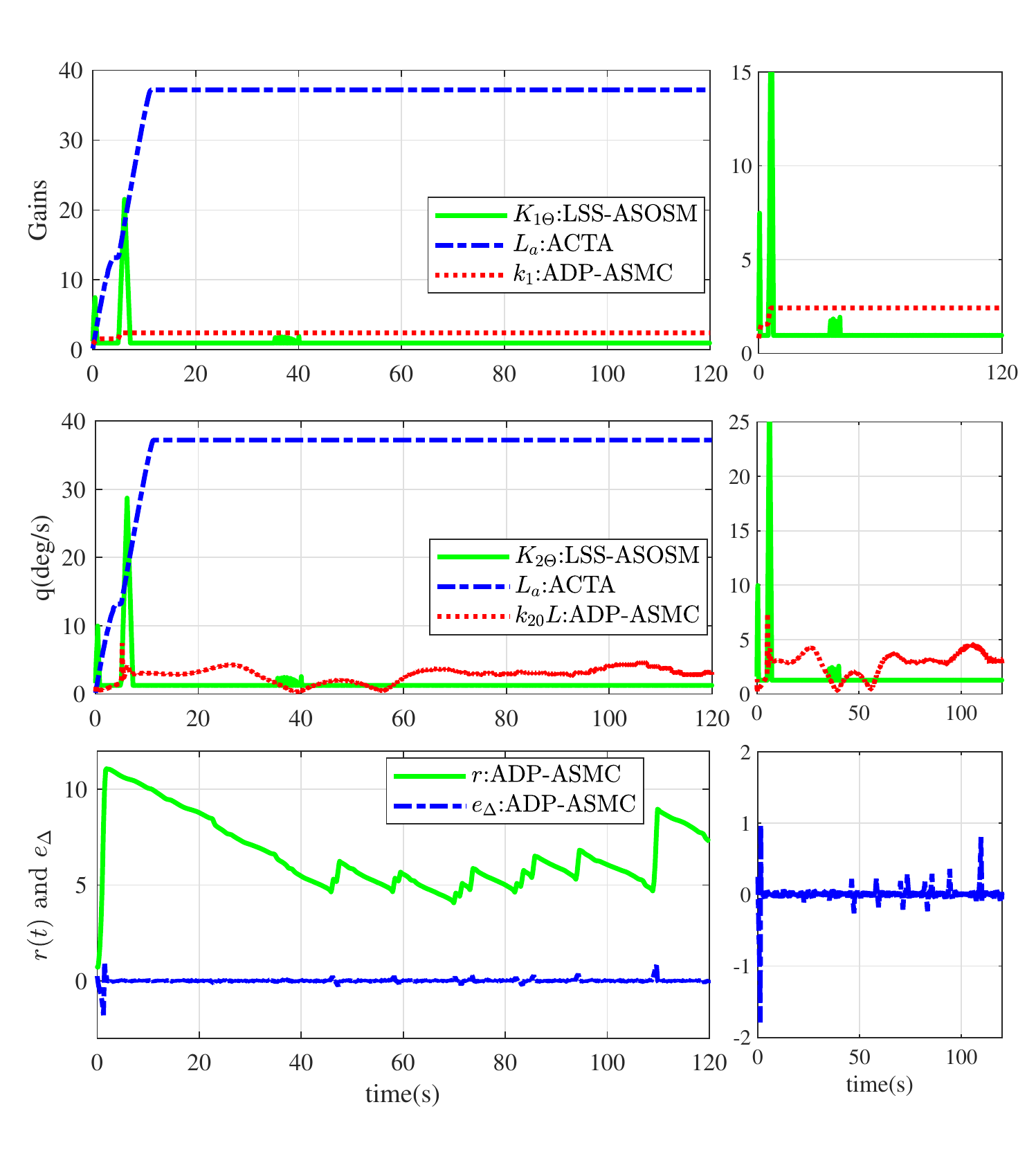}
	\caption{Control gains of LSS-ASOSM, ACTA, and ADP-ASMC.}
	\label{zitaizengyi}
\end{figure}

Fig. \ref{zitaizengyi} shows the control gains.
We can observe that the control gain $L_a$ in ACTA is monotonically increasing and is much larger than that in LSS-ASOSM and ADP-ASMC. Then, control moments with aggressive chattering are generated (shown in Fig.~\ref{liju}). In proposed ADP-ASMC, the control gain $k_{20}L$ changes with the disturbance.
From Fig. \ref{zitaizengyi}, although the control gains $K_{1\Theta}$ and $K_{2\Theta}$ are smaller than $k_1$ and $k_{20}L$, respectively, the convergence precision of tracking errors under LSS-ASOSM is lower than that under the other two control methods. 

\begin{figure}[!htb]
	\centering
	\includegraphics[width=7.8cm]{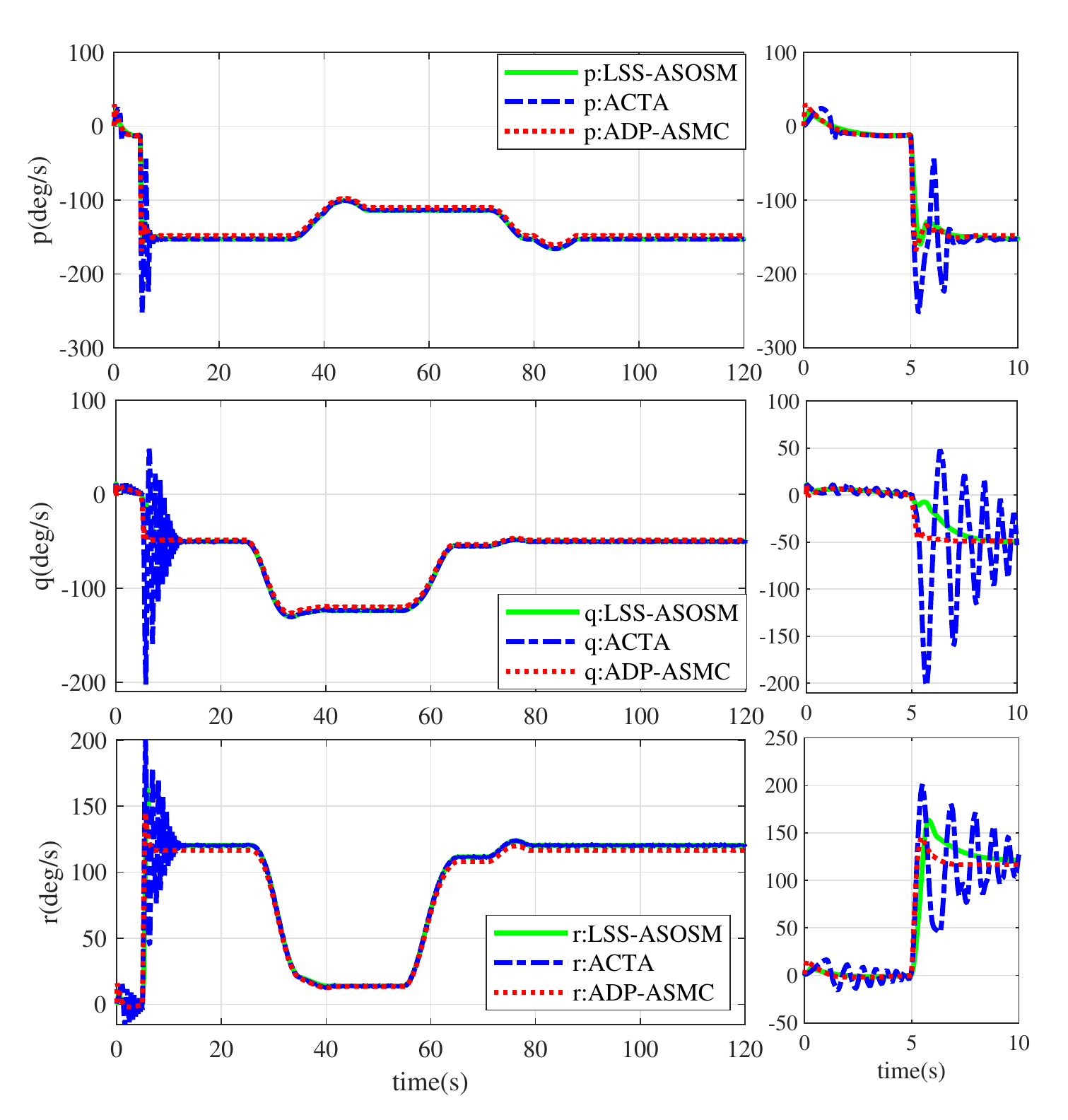}
	\caption{Angular rates $p$, $q$, $r$ of LSS-ASOSM, ACTA, and ADP-ASMC.}
	\label{jiaosulv}
\end{figure}

The variations of the attitude angular rates including $p$, $q$, and $r$ under three methods are shown in Fig.~\ref{jiaosulv}.
It is found that the angular rates of ADP-ASMC change with smaller chattering and faster convergence compared with the other two control methods. 

\begin{figure}[!htp]
	\centering
	\includegraphics[width=7.8cm]{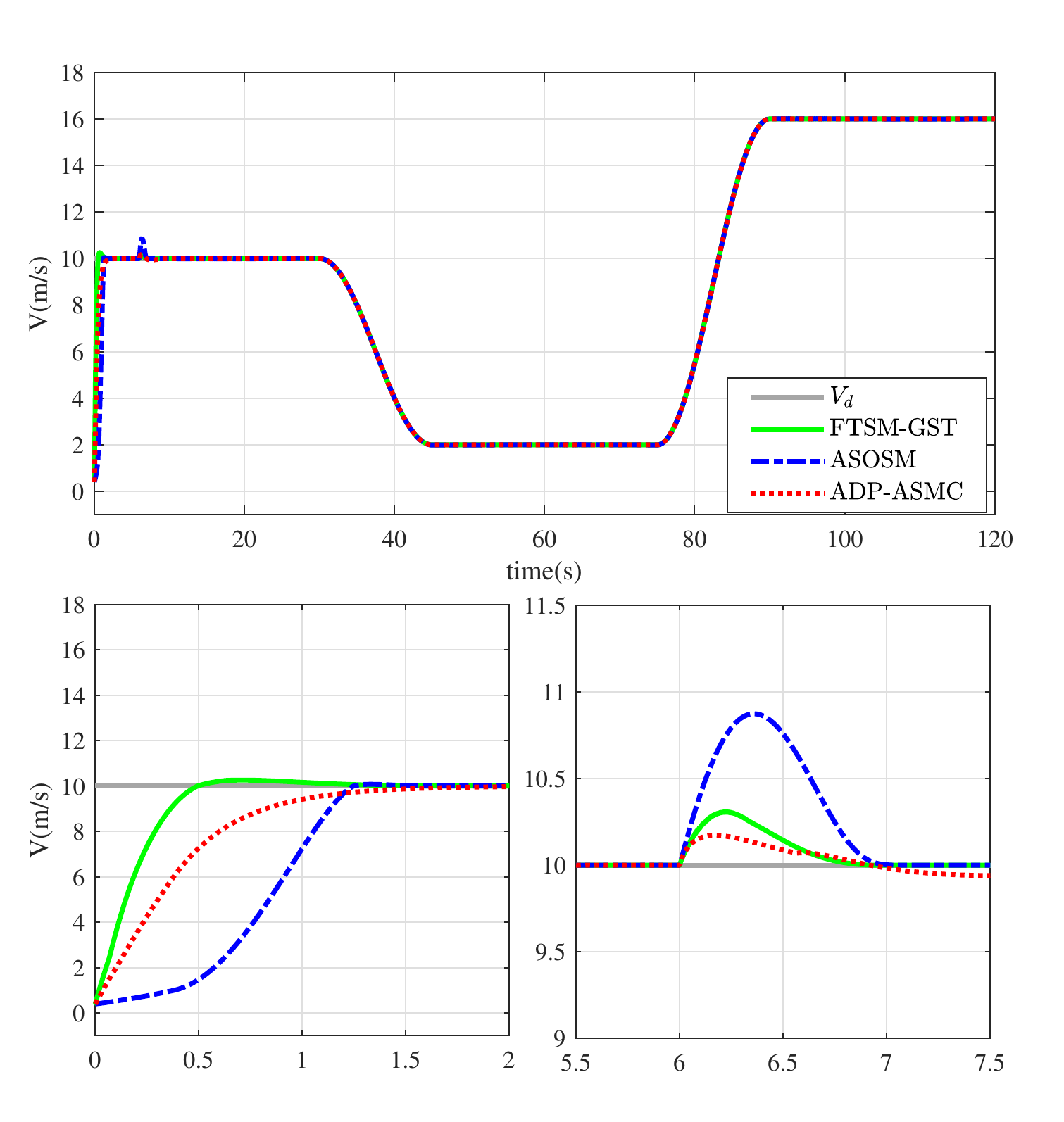}
	\caption{Tracking performance of airspeed subsystem under FTSM-GST, ASOSM, and ADP-ASMC.}
	\label{sudugenzong}
\end{figure}

\begin{figure}[!htb]
	\centering
	\includegraphics[width=7.8cm]{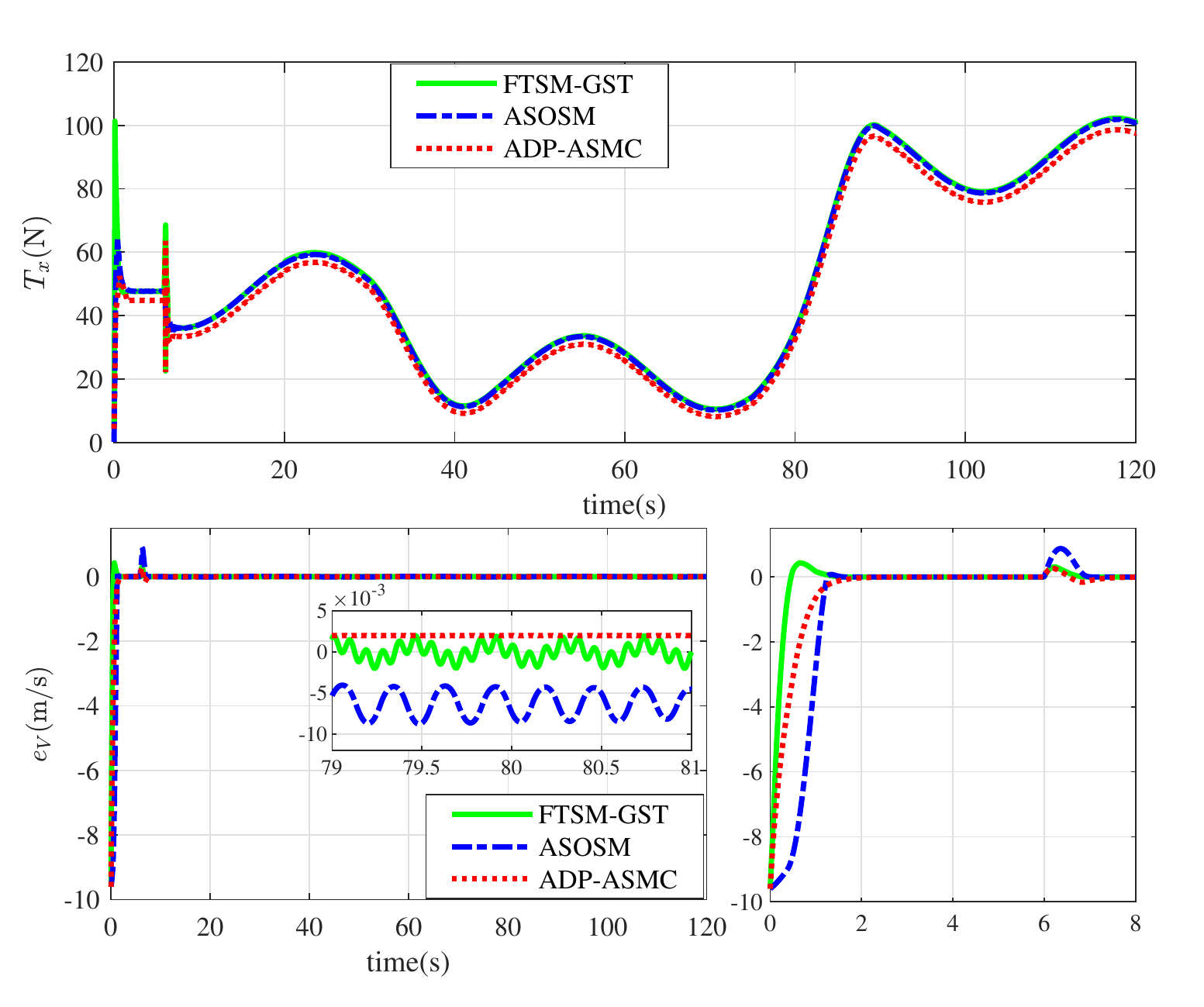}
	\caption{Thrust $T_x$ and tracking error $e_V$.}
	\label{tuilihewucha}
\end{figure}

\begin{figure}[!htb]
	\centering
	\includegraphics[width=7.8cm]{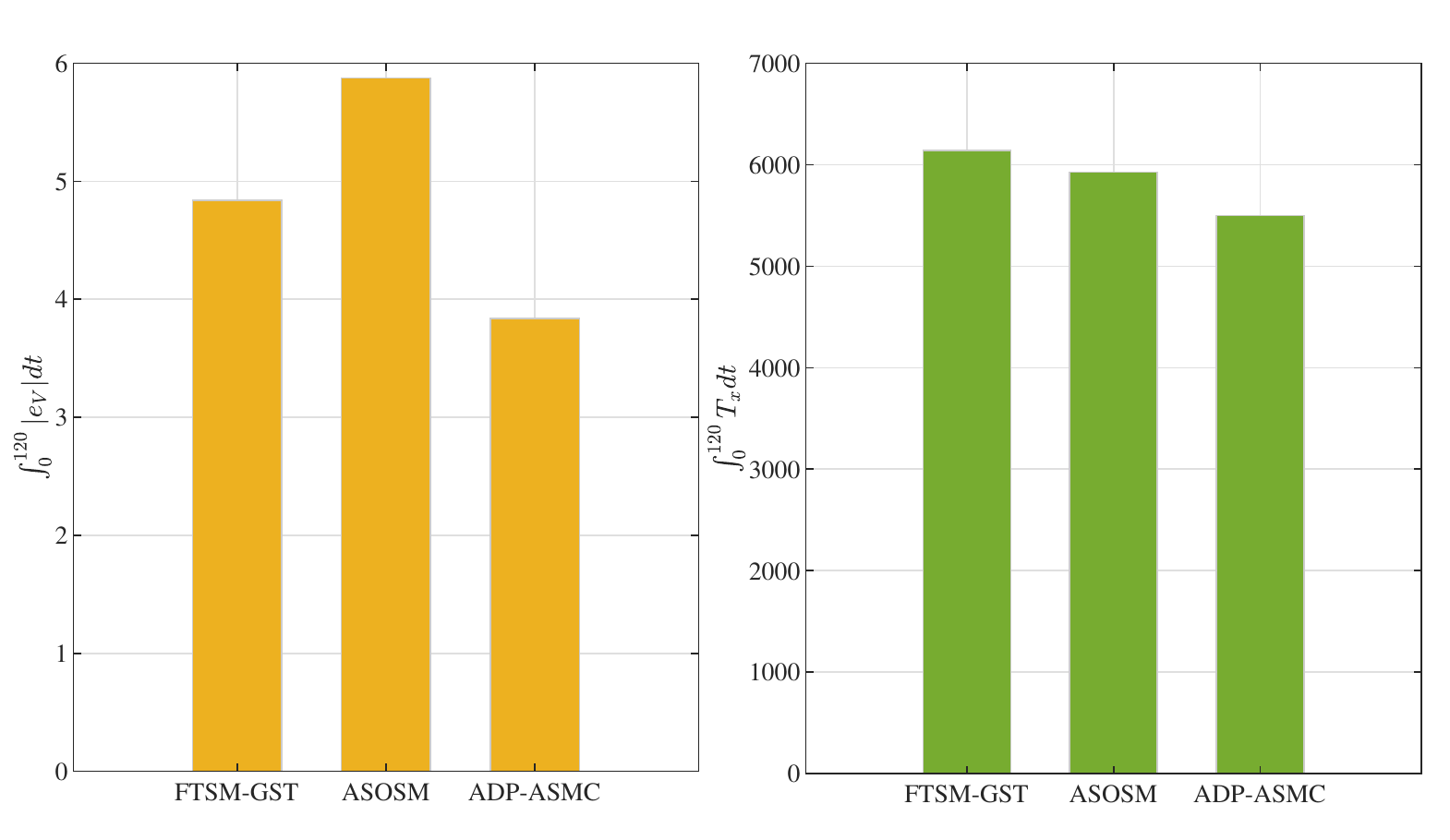}
	\caption{Performance index comparisons of three control schemes.}
	\label{suduperformance}
\end{figure}

The tracking performance of the airspeed subsystem under ASOSM, FTSM-GST and ADP-ASMC are presented in Fig. \ref{sudugenzong}. The airspeed $V$ can track the reference command $V_d$ under three control methods. In comparison with the other two control methods, the proposed ADP-ASMC achieves smaller overshoot. Moreover, the thrust $T_x$ and tracking error $e_V$ are depicted in Fig. \ref{tuilihewucha}. In the top sub-graph of Fig.~\ref{tuilihewucha}, the thrust $T_x$ generated by ADP-ASMC is smaller than that of the other two control methods.  
Besides, similar to the attitude subsystem, $\int_{0}^{120}\left|e_V\right|dt$ and $\int_{0}^{120}T_x dt$ are selected as performance indexes to evaluate the control performances of three control methods in airspeed subsystem. 
The performance index comparisons among FTSM-GST, ASOSM, and ADP-ASMC are shown in Fig.~\ref{suduperformance}.
Through the simulation results in Figs.~\ref{sudugenzong}-\ref{suduperformance}, it is explicit that the proposed ADP-ASMC achieves better dynamic characteristics and lower energy consumption.

\begin{figure}[!htb]
	\centering
	\includegraphics[width=7.8cm]{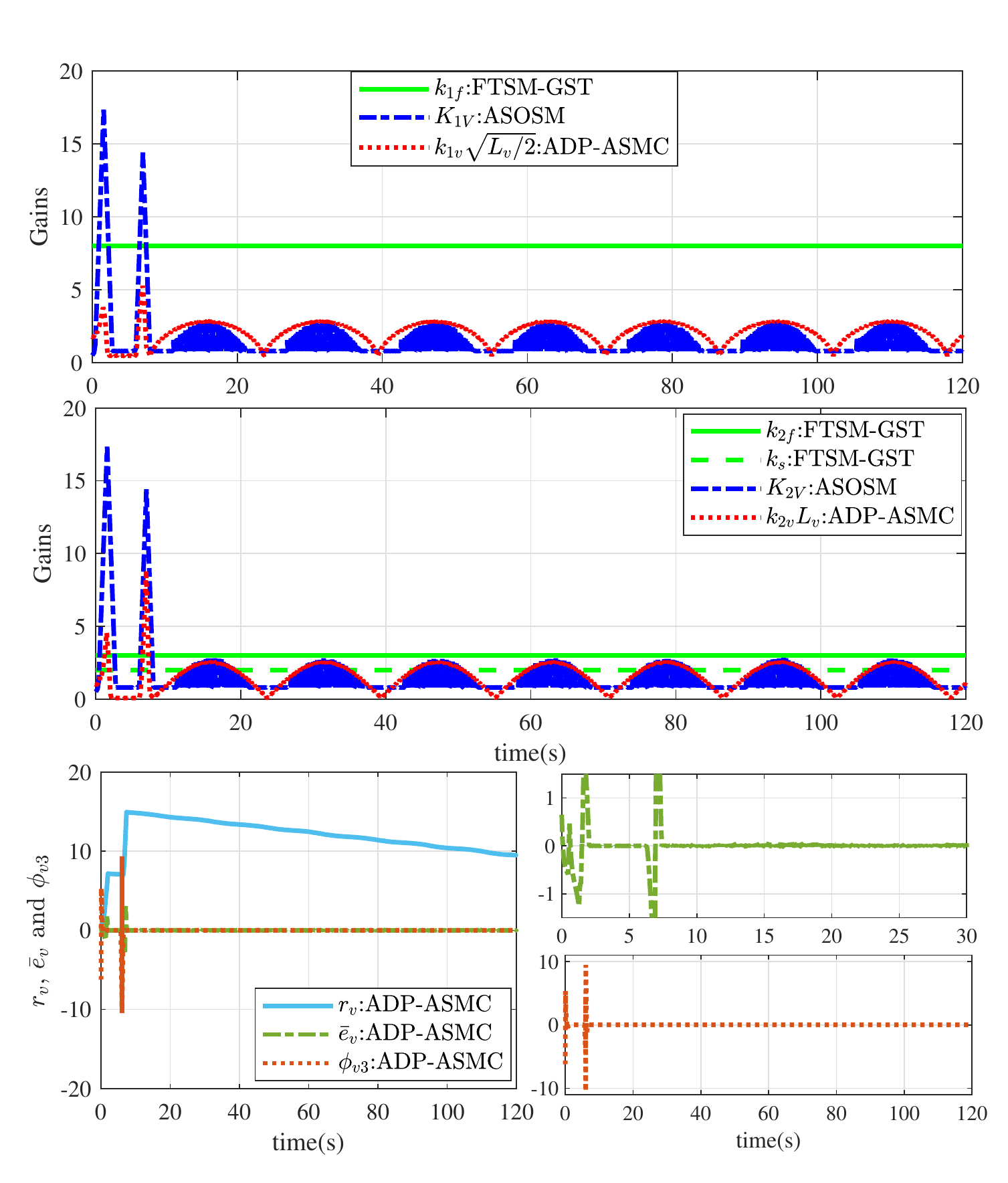}
	\caption{Control gains of FTSM-GST, ASOSM and ADP-ASMC.}
	\label{suduzengyi}
\end{figure}

Fig.~\ref{suduzengyi} shows the control gains in three control methods. After $t=6s$, the control gains in ASOSM and ADP-ASMC are smaller than that in FTSM-GST. However, the convergence precision of ASOSM is lower than that of the other two methods (see the local curves of $e_V$ in Fig.~\ref{tuilihewucha}). In the bottom sub-graph of Fig. \ref{suduzengyi}, $r_v$ and $\bar{e}_v$ are also presented. $\bar{e}_v$ is limited in a finite domain and
$r_v$ starts decreasing after 9s.

\begin{figure}[!htb]
	\centering
	\includegraphics[width=7.8cm]{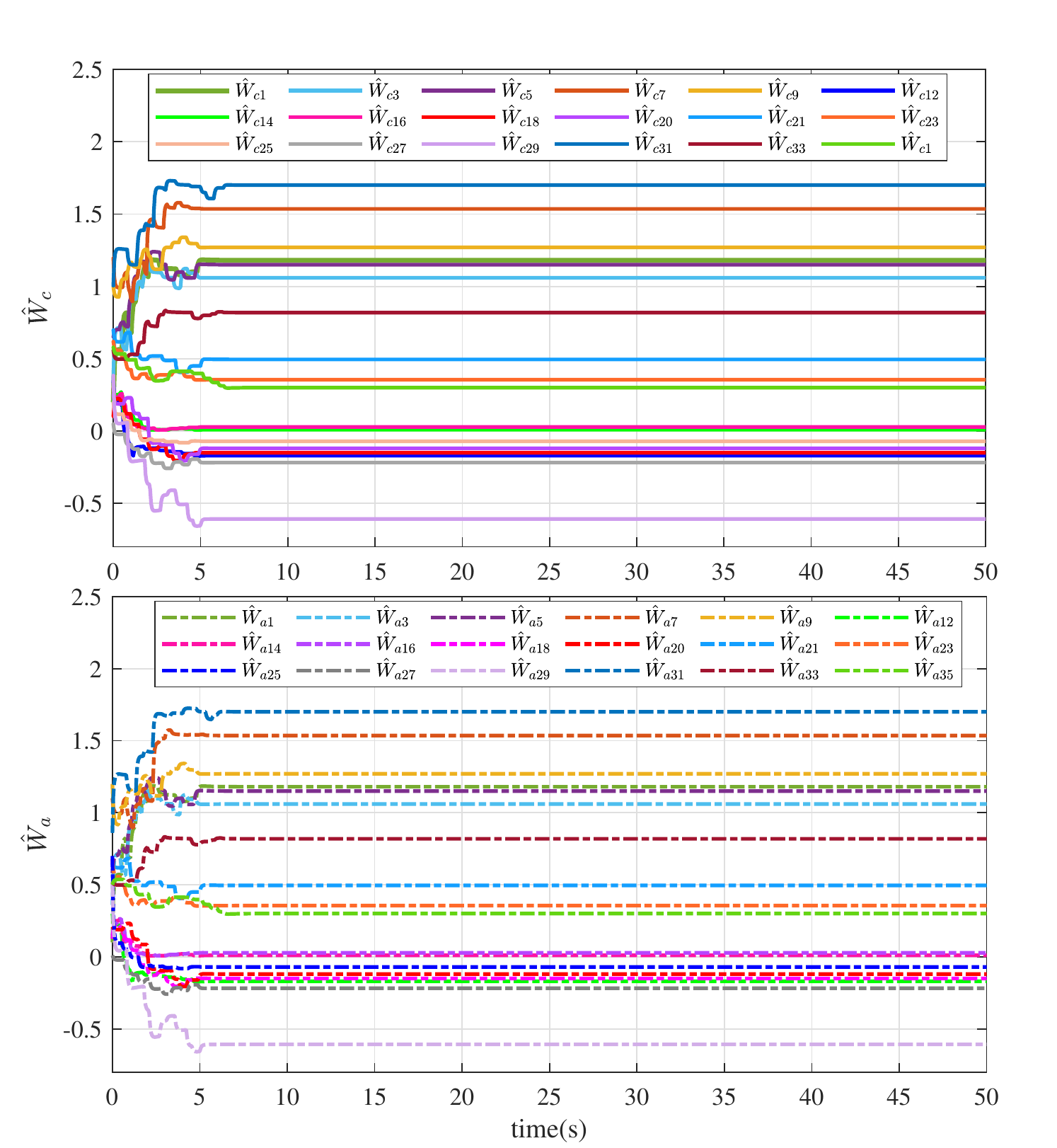}
	\caption{Weight estimations $\hat{W}_c$ and $\hat{W}_a$ of CNN and ANN.}
	\label{weight}
\end{figure}

Fig. \ref{weight} depicts the evolution of CNN and ANN weight vectors $\hat{\bm W}_c$ and $\hat{\bm W}_a$. As indicated in Fig. \ref{weight}, the weights are dramatically adapting during the early stage and the weight vectors of CNN and ANN are both convergent after a few seconds.

\section{Conclusion}
\label{section7}
In our work, the adaptive dynamic programming-based adaptive-gain sliding mode control (ADP-ASMC) scheme is proposed for fixed-wing UAV subject to unknown disturbances.
Aiming at the different issues in two subsystems, the ISM-based AMGST and ISM-based AGST, which are designed via two novel AGST algorithms, are utilized to reject the unknown disturbances such that the equivalent sliding-mode dynamics can be obtained. Unlike the existing adaptive-gain sliding mode algorithms, the two gains in the novel AMGST are tuned through two different gain-adaptation laws so as to address the state- and time-dependent disturbance better and attenuate chattering efficiently. 
Then, a modified ADP approach is constructed to generate the nearly optimal control laws for the sliding-mode dynamics. In this approach, the requirement of initial stabilizing control is relaxed. 
Compared with the conventional sliding mode controllers, the proposed ADP-ASMC scheme provides stronger robustness and meanwhile reduces the energy consumption of fixed-wing UAV. 
Simulation results show that the ADP-ASMC scheme can provide better control performance and demonstrates the superiority of the proposed control scheme.

	\bibliographystyle{model6-num-names}

	\bibliography{UAVbibfile}

\end{document}